\documentclass[fleqn,usenatbib]{mnras}

\usepackage{newtxtext,newtxmath}
\usepackage[T1]{fontenc}

\DeclareRobustCommand{\VAN}[3]{#2}
\let\VANthebibliography\thebibliography
\def\thebibliography{\DeclareRobustCommand{\VAN}[3]{##3}\VANthebibliography}

\usepackage{graphicx}	% Including figure files
\usepackage{amsmath}	% Advanced maths commands
\usepackage{amssymb}	% Extra maths symbols
\usepackage{natbib}
\usepackage{subfig}
\usepackage{booktabs}
\usepackage{upgreek}
\usepackage{setspace}
\usepackage{gensymb}
\usepackage{comment} 
\usepackage{multirow}
\usepackage{changepage}
\usepackage{float}
\usepackage{appendix}
\usepackage{tabularx}
\usepackage{longtable}
\usepackage{epsfig}
\usepackage{morefloats}
\title[Two New Nova Shells]{Two New Nova Shells associated with V4362 Sagittarii and DO Aquilae}

\author[E. J. Harvey et al.]{
E.J. Harvey$^{1}$\thanks{E-mail: e.j.harvey@ljmu.ac.uk}
M.P. Redman$^{2}$ 
P. Boumis$^{3}$ 
S. Akras, $^{4}$ 
K. Fitzgerald,$^{2}$
S. Dulaimi,$^{2}$ 
\newauthor
S.C. Williams,$^{7,8,9}$
M.J. Darnley,$^{1}$  
M.C. Lam,$^{1}$
M. Kopsacheilli,$^{5,6}$  
S. Derlopa,$^{3,10}$ 
          \\
$^{1}$ Astrophysics Research Institute, Liverpool John Moores University, IC2 Liverpool Science Park, Liverpool, L3 5RF, UK \\
$^{2}$ Centre for Astronomy, School of Physics, National University of Ireland Galway, University Road, Galway, Ireland \\
$^{3}$ National Observatory Athens, Inst Astron Astrophys Space Applicat \& Remote Sen, GR-15236 Athens, Greece \\
$^{4}$ Instituto de Matem\'{a}tica, Estat\'{i}stica e F\'{i}sica, Universidade Federal do Rio Grande, Rio Grande 96203-900, Brazil \\
$^{5}$ Physics Department and Institute of Theoretical and Computational Physics, University of Crete, 71003 Heraklion, Crete, Greece \\
$^{6}$ Foundation for Research and Technology-Hellas, 71110 Heraklion, Crete, Greece \\
$^{7}$ Physics Department, Lancaster University, Lancaster, LA1 4YB, UK \\
$^{8}$ Finnish Centre for Astronomy with ESO (FINCA), Quantum, Vesilinnantie 5, University of Turku, 20014 Turku, Finland \\
$^{9}$ Department of Physics and Astronomy, University of Turku, 20014 Turku, Finland \\
$^{10}$ Department of Physics, University of Athens, Athens, Greece
}

\date{Accepted XXX. Received YYY; in original form ZZZ}

\pubyear{2020}

\begin{document}
\label{firstpage}
\pagerange{\pageref{firstpage}--\pageref{lastpage}}
\maketitle

% Abstract of the paper
%\begin{comment}
\begin{abstract}
A classical nova is an eruption on the surface of a white dwarf in an accreting binary system. The material ejected from the white dwarf surface generally forms an axisymmetric shell. 
The shaping mechanisms of nova shells are probes of the processes that take place at energy scales between 
planetary nebulae and supernova remnants. 
%However, the three-dimensional structure of these shells is difficult to untangle when viewed on the plane of the sky. 
%We present two previously undiscovered shells around known classical nova systems, using multi-epoch imaging and spectroscopic analysis, and place them in context with the known nova shell population.
%Imaging and spectroscopic observations of two previously undiscovered nova shells 
%associated with slow nova light-curves are presented. From new narrow-band H$\alpha$+[N~{\sc ii}] 6584$\AA$ and [O~{\sc iii}] 5007$\AA$  observations 
%the better observed nova shell is recreated in the 3D morpho-kinematic code \textsc{shape} and then passed to an adapted version of \textsc{pyCloudy} in order to perform pseudo-3D ionisation simulations.
We report on the discovery of nova shells surrounding the post-nova systems V4362 Sagittarii (1994) and more limited observations of DO Aquilae 
(1925). 
Distance measurements of $0.5\substack{+1.4 \\ -0.2}$\ kpc for V4362 Sgr and 6.7 $\pm$ 3.5 kpc for DO Aql  are found based on the expansion parallax method. The growth rates are measured to be
0.07$^{\prime\prime}$/year for DO Aql and 
0.32$^{\prime\prime}$/year for V4362 Sgr. A preliminary investigation into the ionisation structure of the nova shell associated with V4362 Sgr is presented.
The observed ionisation structure of nova shells depends strongly on their morphology and the orientation
of the central component towards the observer. 
X-ray, IR and UV observations as well as optical integral field unit spectroscopy are required to better understand these interesting objects. 
%\end{comment}
\end{abstract}

%\end{comment}
%%%%%%%%%%%%%%%%%%%%%%%%%%%%%%%%%%%%%%%%%%%%%%%%%%

%%%%%%%%%%%%%%%%% BODY OF PAPER %%%%%%%%%%%%%%%%%%

\section{Introduction}
Classical novae are a sub-type of cataclysmic variable and are characterised by eruption light curves whose progression are observed from radio through to $\gamma$-ray wavelengths. 
Novae are characterised by their optical eruption spectra and light curves. \cite{93lc} classify 
a variety of nova eruption light curves and give physical explanations for many of their features. Unfortunately, these systems do not attract much attention during their quiescent state, however, their shells are probes for many interesting astrophysical processes. Including the degree of clumping as related to shocks in the evolving ejecta shortly post-maximum as well as nebular abundances, which are in turn related to the material accreted before eruption. As there are few nova shells whose structure is resolvable, the discovery of any additional shells allows us to view the population at different ages and investigate their physical properties with the international astronomical community's current ground and space based observational capabilities.

In many cases, the inclination angle has only been constrained by whether the inner binary system does or does not eclipse. As the orientation of nova shells in the plane of the sky is related to that of the binary nucleus \citep{porterasphericity}, estimates of the binary's orbital characteristics can be reached if the geometry of the shell can be untangled. \cite{hutchingHRDEL} was the first to show that the most likely structure of nova shells consisted of an equatorial waist with polar cones of emission. Although there had been previous discussions on how a nova shell's morphology could be derived from observed emission line structure, with early work summarised in \cite{payne}. The effort has been continued in more recent years in \cite{shapeRSoph}, \cite{Ribeiro2011},
\cite{shape_novamon},
\cite{shape_munari_oph},
\cite{shape_kteri}, \cite{GKme}, \cite{me_V5668Sgr} and \cite{2020MNRAS.495.2075P}.

\begin{table*}
\centering
\caption[Shell and binary inclination]{Demonstration of shell and binary orbital inclination dependence. 
Values obtained from the literature, apart from the shell inclinations for GK Per, AT Cnc and Z Cam which were derived during the preparation of \cite{GKme,eamonnthesis}. In this table CN stands for classical nova, DN dwarf nova, RN recurrent nova and PN represents a planetary nebula associated with the listed object. 
The references are as follows: GK Per; \cite{bode87,Morales-Rueda:2002aa}, AT Cnc; \cite{Shara:2012ac}, Z Cam; \cite{Shara:2012ab}, 
V458 Vul; \cite{Wesson458,V458vul_hybrid}, HR Del; \cite{harmanbrien}, DQ Her; \cite{Vaytet}, Nova Mon 2012; \cite{shape_novamon}, 
RS Oph; \cite{shapeRSoph}, T Pyx; \cite{Chesneau:2011aa}, Hen 2-428; \cite{santan-garc}, Hen 2-11; Jones et al. 2014, A\&A, Vol 562, pp 89, 
HaTr 4 \cite{tyndall2012}, Sp1; \cite{jones2012}, Abell 65; \cite{Huckvale:2013}}
\label{inclination_dependence}

\begin{tabular}{lllll}
\toprule
Object		& Type	& Inc. shell & inc. Binary & P$_{orb}$ (days) \\
\\
\hline
\\
GK Per        & CN \& DN \& PN & 54$\pm$5$^\circ$     & 50 - 73$^\circ$                 & 2      \\
AT Cnc        & CN \& DN       & 48$\pm$4$^\circ$  & 17$\pm$3$^\circ$ or 36$\pm$12$^\circ$          & 0.24   \\
Z Cam         & CN \& DN       & 64$\pm$8$^\circ$     & 52 - 69$^\circ$                 & 0.29   \\
%V728 Sco      & CN \& DN       & -          & $\sim$ 82$^\circ$               & 0.138  \\
V458 Vul      & CN \& PN       & $\pm$30$^\circ$       & $\sim$ 30$^\circ$               & 0.068  \\
HR Del        & CN             & 35 $\pm$ 3$^\circ$    & 41$\pm$4$^\circ$                    & 0.17   \\
DQ Her        & CN             & 86.8$\pm$0.2$^\circ$  & 89.6$\pm$0.1                 & 0.19   \\
Nova Mon 2012 & CN             & 82 $\pm$ 6$^\circ$    & ``high inc"       & 0.296  \\
RS Oph        & RN             & 39$\pm$9$^\circ$     & $\sim$30-40$^\circ$             & 455.72 \\
T Pyx         & RN             & $\sim$15$^\circ$  & 10$\pm$2$^\circ$                    & 0.076  \\
Hen 2-428     & PN             & 68$^\circ$        & 64.7$^\circ$                    & 0.175  \\
Hen 2-11      & PN             & $\sim$90$^\circ$  & 90$\pm$0.5$^\circ$                  & 0.609  \\
HaTr 4        & PN             & 65 - 80$^\circ$   & 55 - 75$^\circ$                 & 1.74   \\
Sp 1          & PN             & 10 - 15$^\circ$   & 15 - 25$^\circ$                 & 2.9    \\
Abell 65      & PN             & 68 $\pm$ 10$^\circ$   & 68$\pm$2$^\circ$                    & 1      \\
%Abell 41      & PN             & 66 $\pm$ 5$^\circ$    & 65.7$\pm$0.9$^\circ$                & 0.113  \\
%NGC 6337      & PN             & $<$10 or 15$^\circ$ & 6 $<$ i $<$ 20$^\circ$              & 2.9    \\
%NGC 6778      & PN             & $\sim$ 12$^\circ$ & Models show 12$^\circ$ poss & 0.15  \\
\hline
\label{incsrel}
\end{tabular}
\end{table*}

In the examples of T Aur \citep{gallagher1980}, HR Del \citep{Duerbeck,hutchingHRDEL,HRDel3d}, DQ Her \citep{williamsdqher,Vaytet}, V1500 Cyg \citep{beckerv1500cyg,hutchingsmccall}, V476 Cyg \citep{Duerbeck}, FH Ser \citep{gillbrien}, CP Pup \citep{Duerbeck}, RR Pic \citep{gill}, and GK Per \citep{Liimets:2012aa,GKme}, it is evident that their equatorial waists and polar cone/blob shells have become clumpy. 
The geometry is often complex due to several processes at work. Clumping of the ejecta is likely due to the Richtmeyer-Meshkov instability \citep{Torasker} and is important for the formation of dust in the shells of novae \citep{joiner}. See \cite{spectrophot,Mason_2018} for a framework to unify nova observations in the context of a bipolar shell geometry. Of the resolved nova shell population RS Oph \citep{shapeRSoph}, T Pyx \citep{Chesneau:2011aa} and V1280 Sco \citep{Chesneau:2012aa} demonstrate convincing bipolarity, without discernible equatorial waists.
% RS Oph, T Pyx, V1280 Sco\cite{}

Clumps in equatorial and polar structures are the most probable birth places of carbon and oxygen rich grains, see for example \cite{2018ApJ...858...78G}. An explanation for the existence of tropical rings and polar cones is given within the hydrodynamical work of \cite{porterasphericity} where the tropical rings form by 
sweeping up conical regions of enhanced density local to the matter ejected by the white dwarf.

In the summary of \cite{slavin} several interesting conclusions are laid out that are still relevant. (i) There is a correlation between remnant shape and speed class and (ii) the orientation of the equatorial rings can be used to determine the orbital inclination of nova systems, see Table \ref{incsrel}. 

Shells around classical novae have been searched for and presented in three major published articles: \cite{cohen,gill} and \cite{downes}. The success rate of these searches were 8/17, 4/17 and 13/30 nova shells found around potential candidates, these comprise roughly half of the known nova shells, the other half, for the most part, have been uncovered individually. More recently \cite{absenceofshells} searched for nova shells around nova-like cataclysmic variable systems, without the successful detection of shells around the 15 objects in their study. The non-detection of nova shells around these objects are used to place constraints on the recurrence timescale of the objects in their study. Elsewhere, \cite{sahman} searched the IPHAS archives for nova shells around 101 cataclysmic variable systems, of which three showed evidence of previously unknown associated nebulosity.  

This paper follows the layout described here. First, observations are presented in Section \ref{Observations}. Following this, our analysis and results are presented in Section \ref{anres}. The first two sections follow an internal order of imaging and then spectroscopy. In the discussion section (see Section \ref{discussion}) we look at the how the data presented can be incorporated in to what is known in nova theory and the implications for simulations. Here a report on preliminary ionisation analysis using an adapted version of $\textsc{pyCloudy}$ \citep{pycludy} \footnote{\url{https://github.com/Morisset/pyCloudy}} is also discussed.

\begin{table}
\centering
\caption{WISE magnitudes and derived flux following~\protect\cite{WISE} for a red dominated source. The flux / mag measurement in WISE band 2 corresponds to a S/N of 1.4 and can therefore be used only as an upper limit, the remaining band observations are well sampled. The strong rise red-wards is indicative of either the presence of a cooling dust shell or strong line emission, as would be expected from a coronal nova, see~\protect\cite{wisenova}.}
\label{tab:WISE_V4362sgr}
\begin{tabular}{lllll}
\toprule
Band & 1 & 2 & 3 & 4 \\
\midrule
CWL ($\mu$m) & 3.4 & 4.6 & 12 & 22 \\
WISE mag & 15.117 & 16.105 & 8.664 & 5.501 \\
mag err & 0.274 & 0.3 & 0.029 & 0.043 \\
Flux (Jy) & 0.00027 & 0.000061 & 0.0099 & 0.048 \\
\hline
\end{tabular}
\end{table}

\begin{table*}
\centering
%[htbp]
\caption{Imaging Observations. All images were acquired using the Aristarchos telescope, except for the 2002 Skinakas imaging observations of V4362 Sgr (PTB 42). The column titled t-t$~\protect_{max}$ (yrs) shows the time since nova maximum in years with respect to the observation date. Imaging data described below the double line corresponds to known nova producing systems without discovered shells in this survey. Magnitudes at maximum and minimum are taken from the CBAT list of galactic novae ~\protect\citep{cbatlistnovae}, whose references are given as discovery announcements: where AN = Astronomische Nachrichten, I = IAU Circulars.}
%\begin{tabular}[width=0.3\textwidth]{lc c c c cl}
 %\begin{centre}
\begin{tabular}[width=\protect\pagewidth]{lcclllccc}
\toprule
Obs Date & t-t$~\protect_{max}$ (yrs) &Object & m$~\protect_{max}$ &
m$~\protect_{min}$ & Ref & Filter
CWL/FW(~\protect\AA) & seeing 
($~\protect^{\prime\prime}$) & 
Exp. (sec) \\
\midrule
2018-8-5 & 24.22 &V4362 Sgr (1994) & 3.5? & 15.5 & I5993 & V (RISE2) & 1.6 & 35 \\
2016-8-2 & 22.21 &V4362 Sgr (1994) &  &  &  & H$\alpha$+[N~{\sc ii}] 6578/40 & 1.3 & 2400 \\
 & 22.21 &V4362 Sgr (1994) &  &  &  & 5011/30 & 1.2 & 2400 \\
 & 22.21 &V4362 Sgr (1994) &  &  &  & R 6680/100 & 1.8 & 300 \\
 & 22.21 &V4362 Sgr (1994) &  &  &  & B 5700/70 & 2.2 & 300 \\
2002-5-21 & 8.02 &V4362 Sgr (1994) &  &  &  & H$\alpha$ & 1.6 & 1800 \\
 & 8.02 & V4362 Sgr (1994) &  &  &  & R & 1.8 & $180\times2$ \\
2017-7-24 & 91.85 &DO Aql (1925) & 8.7 & 16.5 & AN225 & H$\alpha$+[N~{\sc ii}] 6578/40 & 2.3 & 2400 \\
 & 91.85 &DO Aql (1925) &  &  &  & R 6680/100 & 2.5 & 180 \\
2015-8-19 & 89.93 &DO Aql (1925) &  &  &  & H$\alpha$+[N~{\sc ii}] 6578/40 & 1.8 & 2400 \\
 & 89.93 & DO Aql (1925) &  &  &  & R 6680/100 & 2.1 & 180 \\
\hline
\hline
2017-7-24 & 118.26 &V606 Aql (1899) & 5.5 & 17.3 & AN153 & [O~{\sc iii}] 5011/30 & 2.6 & 2400 \\

2014-7-20 & 115.25 &V606 Aql (1899) &  &  &  & H$\alpha$+[N~{\sc ii}] 6578/40 & 1.6 & 2400 \\
2014-7-20 & 115.25 &V606 Aql (1899) &  &  &  & R 6680/100 & 1.8 & 180 \\
2016-9-4 & 79.96 &V356 Aql (1936) & 7.7 & 17.7 & I616 & H$\alpha$+[N~{\sc ii}] 6578/40 & 2.4 & 1800 \\
2015-11-18 & 112.67 &DM Gem (1903) & 4.8 & 16.7 & AN161 & H$\alpha$+[N~{\sc ii}] 6578/40 & 1.5 & 1800 \\
2015-11-18 & 112.67 &DM Gem (1903) &  &  &  & R 6680/100 & 1.7 & 180 \\
2015-11-18 & 97.79 &GI Mon (1918) & 5.2 & 18 & AN206 & H$\alpha$+[N~{\sc ii}] 6578/40 & 1.4 & 1800 \\
2015-11-18 & 97.79 &GI Mon (1918) &  &  &  & R 6680/100 & 1.7 & 180 \\
2015-9-14 & 40.27 &V3964 Sgr (1975) & 6 & 17 & I2997 & H$\alpha$+[N~{\sc ii}] 6578/40 & 1.5 & 1800 \\
2015-9-14 & 40.27 &V3964 Sgr (1975) &  &  &  & R 6680/100 & 2.0 & 180 \\
2015-8-19 & 86.24 &BC Cas (1929) & 10.7 & 17.4 & AN243 & H$\alpha$+[N~{\sc ii}] 6578/40 & 1.8 & 2400 \\
2015-8-19 & 86.24 &BC Cas (1929) &  &  &  & R 6680/100 & 2.1 & 180 \\
2015-8-19 & 22.27 &V1419 Aql (1993) & 7.6 & 17 & I5791 & H$\alpha$+[N~{\sc ii}] 6578/40 & 1.6 & 2400 \\
2015-8-19 & 22.27 &V1419 Aql (1993) &  &  &  & R 6680/100 & 2.0 & 180 \\
2014-7-20 & 15.02 &V1493 Aql (1999) & 8.8 & 17.2 & I7223 & H$\alpha$+[N~{\sc ii}] 6578/40 & 1.5 & 2400 \\
2014-7-20 & 15.02 &V1493 Aql (1999) &  &  &  & R 6680/100 & 1.8 & 180 \\
2014-7-20 & 68.90 &V528 Aql (1945) & 7.0 & 18.1 & I1014 & H$\alpha$+[N~{\sc ii}] 6578/40 & 1.6 & 2400 \\
2014-7-20 & 68.90 &V528 Aql (1945) &  &  &  & R 6680/100 & 1.9 & 180 \\
2014-7-20 & 66.08 &V465 Cyg (1948) & 7.3 & 17.0 & I1154 & H$\alpha$+[N~{\sc ii}] 6578/40 & 1.4 & 2400 \\
2014-7-20 & 66.08 &V465 Cyg (1948) &  &  &  & [O~{\sc iii}]  5011/30 & 1.5 & 2400 \\
2014-7-20 & 66.08 &V465 Cyg (1948) &  &  &  & R 6680/100 & 1.8 & 180\\
\hline
\label{observationstab}
\end{tabular}
\end{table*}

\section{Observations}
\label{Observations}

\subsection{Imaging}
\label{Imaging}
\subsubsection{WISE}

As demonstrated for GK Per (see Fig. 13 of \cite{GKme}), classical nova systems with known shells can be seen in the WISE image archive. However, not only the inner shells are captured, but also, thanks to the nature of the survey, the interaction of ejecta from previous nova events with the interstellar medium can be identified. 

\begin{figure*}
%\centering
%\resizebox{{\columnwidth}{!}{
\includegraphics[width=16cm]{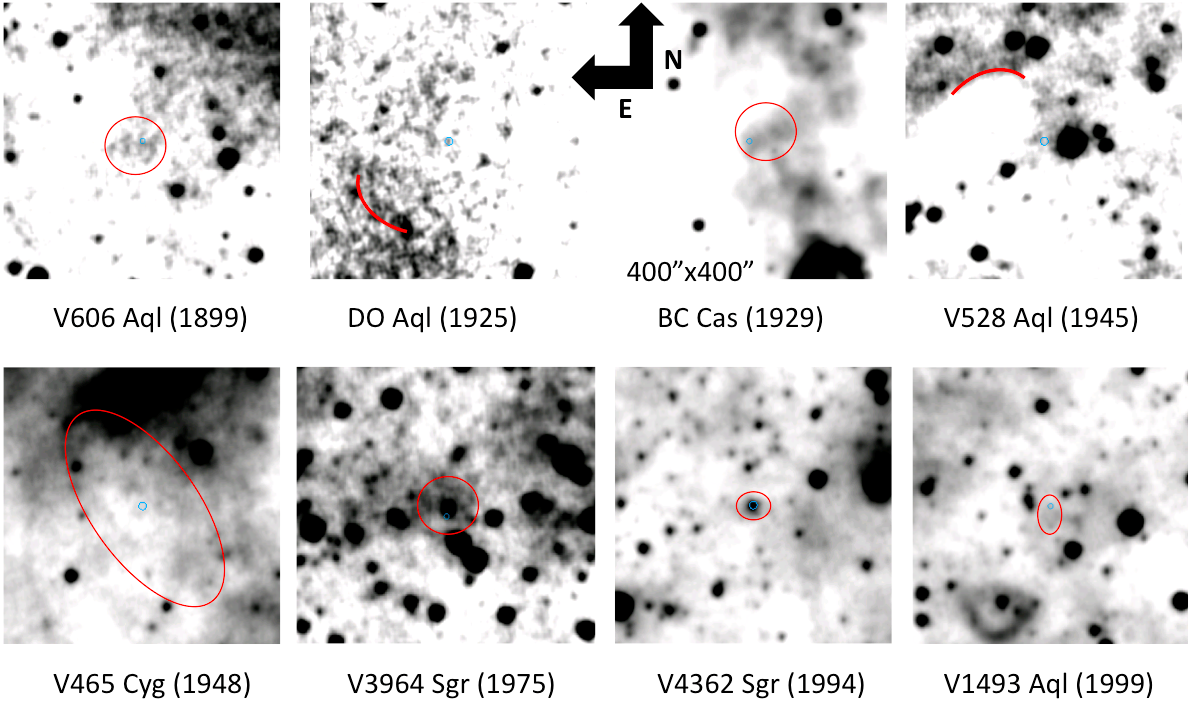}
\caption{Examples of novae that revealed tentative evidence of shells or other interesting features for follow up from the WISE image archive. The displayed images are from WISE band 3. All images are 400"×400", with north up and east to the left. From left to right and top to bottom: V606 Aql displayed interesting emission 0.012${\degree}$ and 0.06${\degree}$ to the SE of the nova progenitor (see Fig. \ref{fig:WISE_V606Aql}); DO Aql a possible shock feature 0.037${\degree}$ to the SE; BC Cas features 0.013${\degree}$ W and 0.032${\degree}$ SW; V528 Aql a feature 0.004${\degree}$ E and a possiblly associated arc 0.037${\degree}$ to the NE; V465 Cyg showed a bright source in WISE bands 1 and 2 and possibly an associated large scale ring in bands 3 and 4 of about 0.03${\degree}$ in radius; V4362 Sgr source is very bright in WISE bands 3 and 4 and V1493 Aql showed possible hints of a nested set of nova shells that would have been associated with previous nova episodes.}
\label{fig:WISEhints}
\end{figure*}

\begin{figure}
%\centering
%\resizebox{{\columnwidth}{!}{
\includegraphics[width=\columnwidth]{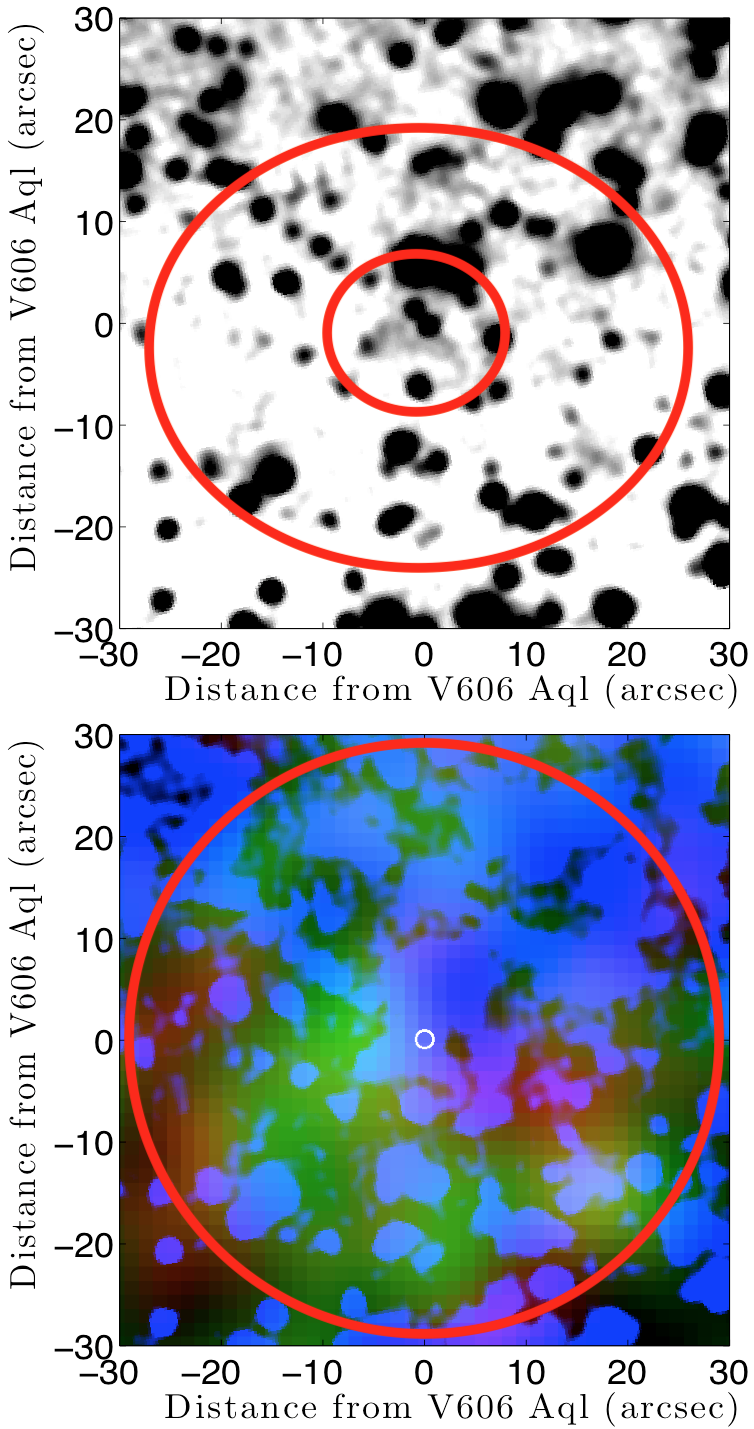}
\caption{V606 Aql (1899) shows tentative evidence of harbouring visible evidence of a shell, albeit not enough to confirm its visibility and has been left for further investigation, with the same being true for BC Cas.  This is largely due to the objects proximity to the Cygnus Rift and the presence of bright neighbouring stars. The top panel shows the Aristarchos H$\alpha$ image, whereas the bottom panel is an overlay of Wise bands 3 (blue) and 4 (green). Both panels are 60"x60" and are centred on the V606 Aql nova progenitor. The red circles mark the suspected nova shell related emission, mentioned in Fig. \ref{fig:WISEhints}.}
\label{fig:WISE_V606Aql}
\end{figure}

Following this a search through the known nova database of novae without previously documented shells was conducted.
The search consisted of acquiring the publicly available multi-band WISE images of reasonably bright nova systems that were observed at nova maximum at least more than 15 years previous to the start of the study (in 2016). The nova systems were found in the CBAT list of novae \protect\citep{cbatlistnovae}. Several of these nova systems showed plausible hints of associated nebulosity in the WISE survey \citep{WISE}, see Figs. \ref{fig:WISEhints} and \ref{fig:WISE_V606Aql}. These hints of nebulosity may have been in the form of bright WISE bands 3 and 4 relative to bands 1 and 2, as was the case for V4362 Sgr, see Table \ref{tab:WISE_V4362sgr}. Or of a suggestion of interaction between previous nova events with the interstellar medium. In the case of the latter, as can be seen marked by red circles and lines in Fig. \ref{fig:WISEhints}, with a corresponding description in the associated caption. Figure \ref{fig:WISE_V606Aql} is a close up of suspected associated emission in the Aristachos and WISE images of the first one of these nova, V606 Aql (1899).

\subsubsection{Aristarchos}

Using the Aristarchos telescope in Greece, deep imaging observations were acquired of the vicinity surrounding the twelve 
classical nova systems in Table \ref{observationstab}. Deconvolved images using MEM and Lucy algorithms were produced for all novae during the survey. This was done along with radial cuts and R band subtraction (difference imaging) for each nova shell candidate, where possible. Unfortunately, with neighbouring stars within 1" for both novae the deconvolved images have artefacts present and the broad band subtracted image versions were deemed the most clear representations.

Of the nova progenitor systems without previously known nova shells 2 of the 12 systems uncovered the unambiguous presence of visible shells, i.e. those presented here. For a list of all novae observed with imaging in this survey see Table \ref{observationstab}. Of the remaining systems V606 Aql and V528 Aql demonstrated faint emission that may be recovered from deeper observations. 
The other eight systems (V356 Aql, V1419 Aql, V1493 Aql, V465 Cyg, BC Cas, DM Gem, GI Mon, and V3964 Sgr) showed no evidence for resovable shells at the time of observation with the instrument setup used (i.e. V1419 Aql and V1493 Aql were not expected to have resolvable shells with the instrumentation used and no broadband filter observation of V356 Aql was acquired due to weather constraints during the night). 

The Aristarchos imaging observations consisted of either one or two narrow-band filters focused on H$~\protect\alpha$
+ [N~{\sc ii}] (6578\protect\AA~/ 40\protect\AA, i.e. central wavelength in~\protect\AA~/ filter width in~\protect\AA) and/or 
[O~{\sc iii}] 
(5011\protect\AA/30\protect\AA) with exposures of 30 or 40 minutes in each filter, 
see Table \ref{observationstab}. For nova systems in quiescence the majority of the continuum 
emission comes from the secondary and the reinstated accretion disk. Imaging in R band was collected 
of each object in order to subtract the stellar continua from the images. 
The seeing during observations was of the order of 1-2$^{\prime\prime}$. The CCD detector has dimensions of $2048\times2048$ pixels 
with each pixel being 24 $\mu$m across
($\approx$ 0.28$^{\prime\prime}$~ per pixel). The imaging data were reduced using standard routines in 
{\sc iraf} \footnote{IRAF is distributed by the National 
Optical Astronomy Observatories, which are operated by the Association of Universities 
for Research in Astronomy, Inc., under cooperative agreement with the National Science Foundation.}.
Again, see Table \ref{observationstab} for a summary of the new imaging discussed in this work.

\subsection{Spectroscopy}
\label{Spectroscopy}

High-resolution echelle spectroscopic data were obtained to 
measure the observable kinematics of the V4362 Sgr nova shell.
These were obtained using the Manchester Echelle Spectrograph (MES) instrument mounted on the 2.1\,m 
telescope at the San Pedro M\'artir (SPM) observatory in Mexico \citep{MeaburnMES}. Instead of using a cross disperser, as in a regular echelle spectrograph, 
an interference filter isolates the desired spectral orders for high resolution observations of nebular lines.
The slit positions were observed with the 
instrumentation in its f/7.5 configuration. A Marconi $2048\times2048$ CCD was used with a resultant spatial 
resolution $\simeq$ 0.35 arcsec pixel$^{-1}$ after $2\times2$ binning was applied during observation with 
a {\raise.17ex\hbox{$\scriptstyle\sim$}}6$'$ long-slit.  Bandwidth filters of 90 and 60\protect\AA~ were 
used to isolate the 87$\protect^{\mathrm{th}}$ and
113$\protect^{\mathrm{th}}$ orders containing the
H$\alpha$+[N~{\sc ii}] 6548\protect\AA, 
6584\protect\AA~and [O~{\sc iii}] 5007\protect\AA~nebular emission lines.  

The nova shells surrounding DO Aql and V4362 Sgr were detected using the low-resolution, high-throughput SPRAT spectrograph \citep{sprat} on the Liverpool Telescope \citep{Lpooltel} during mid 2017 and mid 2018 respectively. The SPRAT observations were taken in blue optimised mode without on-chip binning. With sidereal tracking on and a mount angle of 11$^{\mathrm{o}}$. Although this type of observation is of lower spectral resolution, it still allows to apply velocity constraints and has a broad wavelength coverage, which is necessary for first-pass nebular analysis. 

For a summary of the spectroscopy observations see Table \ref{spectralastchap}. For line flux measurements see Table \ref{linefluxes} and for the calculated line ratios see Table \ref{lineratios2}.

% Please add the following required packages to your document preamble:
% \usepackage[normalem]{ulem}
% \useunder{\uline}{\ul}{}
\begin{table*}
\centering
\caption{Summary of spectroscopy observations undertaken for this work. In this table P.A. stands for position angle of the slit on the plane of the sky. \textit{R} represents spectral resolution quoted in terms of velocity resolution at H$\alpha$. Resolution at [O~{\sc iii}] can be approximately calculated by multiplying the resolution at H$\alpha$ by 0.75. With regards to the MES slit widths 150$\mu$m corresponds to 1.9" on the plane of the sky and 300$\mu$m to 3.8", thus smaller than the measured extent of both recovered nova shells.}
\label{spectralastchap}
\begin{tabular}{llllllll}
\toprule
Object    & Instrument & Filter/Grism  & Slit    & \textit{R} at H$\alpha$ & P.A. ($^{\circ}$) & Exp. (sec) & Date Obs   \\
          &            & CWL/FW(\protect\AA)
          & ($\mu$m) & (km s$^{-1}$) &                   &            &            \\
\hline
V4362 Sgr & MES        &    [O~{\sc iii}] 70\protect\AA            & 150   & 10  & 90                &     1800       & 19/05/2012       \\
V4362 Sgr & MES        &    H${\alpha}$ 90\protect\AA           & 150 & 10    & 90                &       1800     & 19/05/2012       \\
V4362 Sgr & MES         & [O~{\sc iii}] 70\protect\AA      & 300   & 20  & 150               & 1200           &    31/08/2016          \\
V4362 Sgr & MES          & H${\alpha}$ 90\protect\AA        & 300  & 20   & 150               & 1200           &    31/08/2016          \\
V4362 Sgr & MES          & [O~{\sc iii}] 70\protect\AA      & 300  & 20   & 60                &     1200       &      31/08/2016        \\
V4362 Sgr & MES        & H${\alpha}$ 90\protect\AA        & 300  & 20   & 60                &   1200         &     31/08/2016         \\
DO Aql    & SPRAT      &   5827/4685            &   150    & 850  &     0              &  $1200x3$          & 18/06/2017      \\
V4362 Sgr  & SPRAT      &   5827/4685            &   150    & 850  &     0              &  $1200x3$          & 10/07/2018      \\

\hline

\end{tabular}
\end{table*}

%%%%%%%%%%%%%%%%%%%%%%%%%%%%%%%%%%%%%%%%%%%%%%%%%%%

\section{Analysis and Results}\label{anres}

\subsection{DO Aql (1925)}

With a poorly observed eruption light curve and no early spectral observations (i.e. first three months) this system was not recognised initially 
as a nova and was referred to as ``Wolf's Variable" following discovery \citep{DOAqlVorontov}. The system 
was proposed to be a recurrent nova and the star of Bethlehem by \cite{starOBeth}, which was subsequently refuted in 
\cite{doaqlbethleham} based on the recurrence time scale, among other factors. DO Aql is known to have been a slow nova and was thought to have experienced a 
long plateau at maximum of approximately 250 days, with a 53 day gap in observations. The t$_{3}$ (time taken for the nova to decline by three magnitudes from maximum light) of the 
nova event is reported in \cite{doaqlbethleham} to be 900 days, where the visual maximum was reported 
as 8.7 in V. If the maximum was missed it may have occured during the 53 day gap in observations, or else if it occured prior to the discovery date the t$_{3}$ value may then have been derived from a long decline often 
seen in slow novae after a strong dust-dip. If the maximum is indeed as was reported then the DO Aql eruption light curve would be a precursor example of the extremely slow nova V1280 Sco. 
Where V1280 Sco is the slowest nova known to date in terms of early photometric and spectroscopic evolution, as well as the lowest recorded velocity expanding shell  \citep{Chesneau:2012aa}. This would suggest that V1280 Sco occurred on a low mass white dwarf. However, the higher expansion velocity shell of DO Aql does not fit into this comparison.
%Who called it that? Can't get Vorontov paper to load on comp: Early observers of the system described it as a `wave nova' based on the appearance of its early light curve, possibly analogous to the variations observed in the bright recent nova V5668 Sgr, see \cite{me_V5668Sgr}.

The DO Aql system is comprised of an eclipsing 
binary with a period of 4.03 hours, whose quiescent light curve is proposed to demonstrate either 
obscuration of a hot component or stream overshoot \citep{shafter_2neweclipse}. Photometry 
in the BVRJK band-passes are presented in \cite{szkody}, whom observed it to have a V magnitude of 17.66, with colours (B-V) = 0.60 and (V-R) = 0.39 during September 1998.

The observations of this object presented here reveal a previously undiscovered nova shell visible in the 
H$\alpha$+[N~{\sc ii}] narrow band filter image, however [O~{\sc iii}] emission cannot be confirmed from the observations presented here. Two epochs are presented for the H$\alpha$+[N~{\sc ii}] narrow band 
imaging, from 2015 and 2017, see Table \ref{observationstab}, i.e. 90 and 92 years since the observed nova eruption, implying a growth rate of 0.07$^{\prime\prime}$/year. The 2017 observations are affected by poor Seeing conditions, see Table \ref{observationstab}, and as such the 2015 H$\alpha$+[N~{\sc ii}] narrow band image should provide the most accurate distance estimate. However, both epochs of H$\alpha$+[N~{\sc ii}] were used to provide a distance estimate to the shell, see Table \ref{tab:distmeas}.  

\begin{figure*}
%\centering
%\resizebox{{\columnwidth}{!}{
\includegraphics[width=17cm]{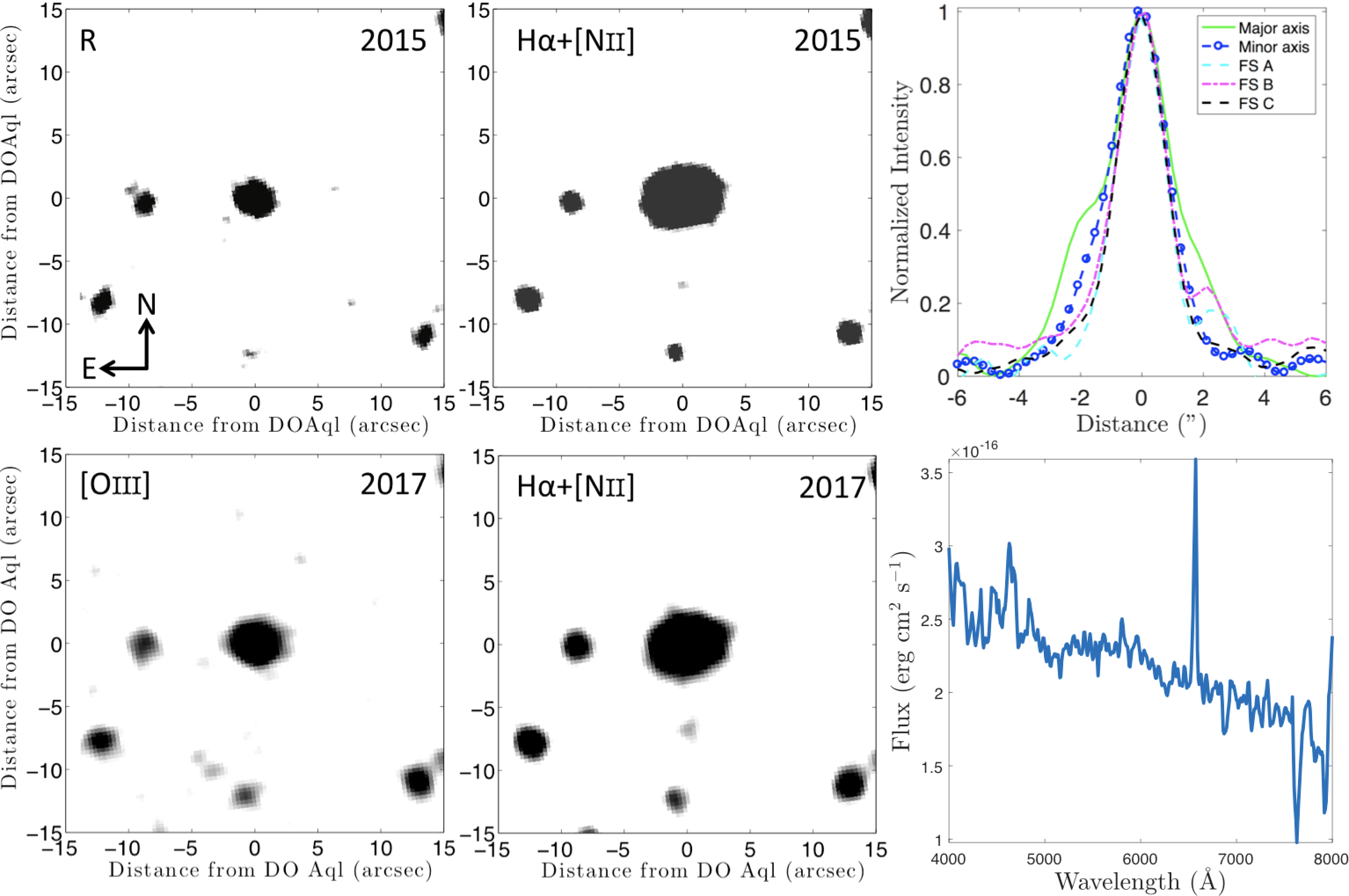}
\caption{Observations of DO Aql (1925). From left to right, top to bottom the panels show: (i) R band image of DO Aql from 2015 (ii) H$\alpha$ + [N~{\sc ii}] 2015 Aristarchos image  
(iii)  Radial cut of DO Aql in comparison to three field stars 
(iv) 2017 [O~{\sc iii}] Aristarchos image (v) 2017 Aristarchos H$\alpha$ + [N~{\sc ii}] image 
(vi) Flux density calibrated SPRAT spectrum of the nova shell and remnant 2017, the features shortward of H$\beta$ are contaminated by noise.}
\label{fig:DOAql}
\end{figure*}

A SPRAT spectrum taken in June 2017 shows the presence of [N~{\sc ii}] emission lines originating 
from the nova shell. The relative extension of the different emissions can be seen in Fig. \ref{fig:DOAql}, this is expected to have an influence on the measured line ratios. The whole nebular spectrum is contaminated by the spatial resolution constraints of the instrument and seeing during the SPRAT spectrum observation, see Fig. \ref{fig:DOAql} and Table \ref{tab:distmeas}.  

As an old and bright nova shell surrounding an eclipsing binary this object is attractive for 
follow-up studies with larger optical telescopes as well as in other wavelength regimes. Such observations would allow to probe the dust properties of the nebula and well as further investigate its chemical, ionisation and physical structure, such as was done for HR Del in \cite{HRDel3d}, T Pyx \citep{shara97,Chesneau:2011aa} and GK Per \citep{Shara:2012aa, Liimets:2012aa, GKme}.

%\begin{landscape}
%\end{landscape}

\begin{table*}
\centering
\caption{Line flux densities, all measurements are $\times10^{-16}$ erg cm$^{-2}$ s$^{-1}$ \protect\AA$^{-1}$. Errors are of the order of 10$\%$ and are not corrected for reddening. $\lq$-' denotes when lines were unresolved and therefore could not be measured.}
\label{linefluxes}
\begin{tabular}{ccccccc}
\toprule
Object    & H$\beta$  & H$\alpha$ & He~{\sc ii} 4686\protect\AA & [O~{\sc iii}] 5007\protect\AA & [N~{\sc ii}] 6548\protect\AA & [N~{\sc ii}] 6583\protect\AA \\
\hline
PTB 42  & 2.3 & 10.4 &  - & 5.6 & 11.2 & 33.1 \\
DO Aql    & 3.4 & 4.5 &  3.7 & - & -  & - \\
\hline
\end{tabular}
\end{table*}

\subsection{V4362 Sgr (1994)} 

V4362 Sgr (1994) was discovered on 1994 May 16.733 UT by Yukio Sakurai and had a maximum observed magnitude of 7.5. Similarly to DO Aql, V4362 Sgr was a poorly observed nova in terms of photometry during the later 
development of its optical light curve, despite being caught on its rise to maximum. However, it was well observed in 
terms of early-time polarimetry, see \cite{evanspol} where complementary photometry of the nova is also presented. 

As a poorly 
observed nova in eruption it is difficult to determine the light curve type although it seems to resemble 
that of DQ Her \citep{93lc}. The DQ Her light curve demonstrated jitters on an otherwise flat top during and shortly after maximum, which was then followed by a dust formation event that was observed via a strong $\lq$dip' in the post-maximum light curve. A spectrum was obtained of the system a week after discovery, described in \cite{Sakurai94} who classified it to be a post-maximum Fe~{\sc ii} type nova. Maximum \textit{observed} light came a month later on 17 April. In support of a missed maximum, the nova would not have been observable a month earlier as it was too close to the Sun. Dust-dip novae often obtain maximum magnitudes 3-4 brighter than what the recovery reaches after a dust-dip, e.g. DQ Her, FH Ser, T Aur, V705 Cas and NQ Vul as decribed in \cite{93lc}. Therefore, the maximum observed visual magnitude could have been around 3.5. This would help explain the small derived distance to the nova, given the implied weak implied absolute magnitude if the nova maximum is attributed to the observed maximum. 

The nova was observed with broadband polarimetry, presented in \cite{evanspol}, that 
covers 51-83 days post discovery where they find that the observed 
absolute polarisation was mostly due to scattering by small dust grains in an axisymmetric shell, possibly consisting of narrow 
conical polar caps and a flattened circular equatorial ring. The proposed structure from the polarimetry in 
\cite{evanspol} is very similar to that proposed for V5668 Sgr in \cite{me_V5668Sgr} from comparable observations, i.e. where both are suggestive of polar caps and a flattened equatorial ring. The ALMA observation of V5668 Sgr in \cite{almaV5668Sgr} are consistent with a shell geometry of polar cones and and equatorial ring, as proposed in \cite{me_V5668Sgr}.
Also, the spectropolarimetric observations of V339 Del in  \citep{kawahita} again demonstrates a similar shell shape.

\begin{figure*}
%\centering
%\resizebox{{\columnwidth}{!}{
\includegraphics[width=17cm]{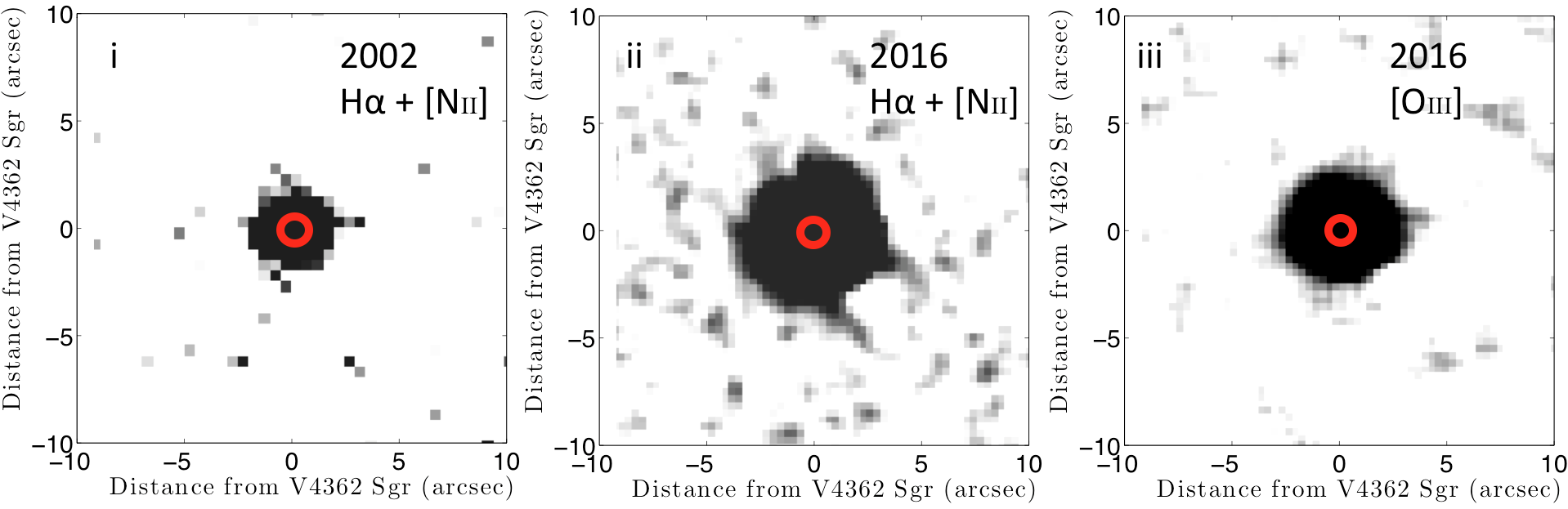}
\caption{Imaging observations of V4362 Sgr/PTB 42: (i) H$\alpha$ + [N~{\sc ii}], Continuum subtracted Skinakas May 2002 image with measured shell dimensions of $2.5^{\prime\prime} \times 3.1^{\prime\prime}$ (ii) H$\alpha$ + [N~{\sc ii}],  Aristarchos 2016 ($6.4^{\prime\prime} \times 7.1 ^{\prime\prime}$ shell size in continuum subtracted image) (iii) [O~{\sc iii}], Aristarchos 2016, ($5.2^{\prime\prime}\times5.5^{\prime\prime}$ measured shell size in the continuum subtracted image). North is up and East is to the left. The red circle on each image shows the seeing disk corresponding to the FWHM of the respective observations.}
\label{fig:ptb42_imag}
\end{figure*}

\begin{figure}
\centering
%\resizebox{{\columnwidth}{!}{
\includegraphics[width=\columnwidth]{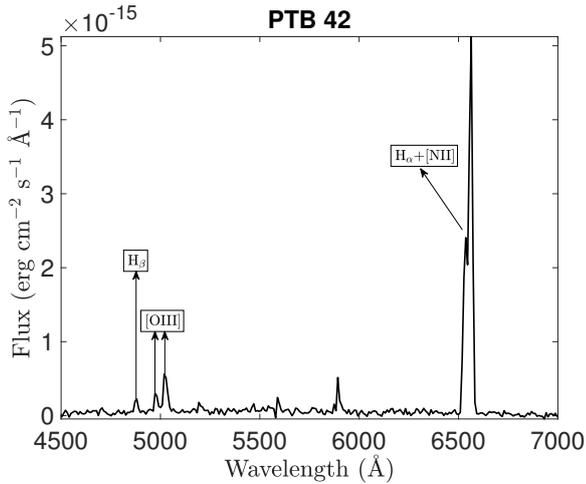}
\caption{Flux calibrated Liverpool Telescope SPRAT spectrum of the PTB 42 shell surrounding the nova position of V4362 Sgr. The feature around 5876\protect\AA~is a sky residual and not He I. The H$\alpha$ emission is blended with the two stronger [N~{\sc ii}] lines.}
\label{fig:spratptb42}
\end{figure}

Here we present a newly discovered nova shell surrounding the nova progenitor V4362 Sgr. As the nova shell was 
observed, but misclassified as a planetary nebula \citep{boumisptb42}, it is possible to present multi-epoch narrow-band 
imaging, see Figs. \ref{fig:ptb42_imag} \& \ref{fig:ptbexpand}. Multi-epoch high-resolution MES spectroscopy was obtained (Fig. \ref{fig:ptb42_spec}), along with a low-resolution SPRAT spectrum (Fig. \ref{fig:spratptb42}).  
The diameter of the nebulosity is recorded as 4$^{\prime\prime}$ in \cite{boumisptb42}. As the nebular object was named PTB 42 in \cite{boumisptb42} that name will be used here to describe the nebular component and V4362 Sgr will refer to the nova progenitor system. After retrieving the original 2002 imaging data 
and subsequent subtraction of the stellar contribution gave the following measured values of the Crete 1.3m Skinakas telescope H$\alpha$ + [N~{\sc ii}] 
imaging data (described first in \cite{boumisptb42}): minor axis = 2.5$^{\prime\prime}$, major = 3.1$^{\prime\prime}$, an axial ratio of 1.2, see Tables \ref{tab:distmeas} and \ref{Inputparametersformodels}. The spectrum presented of PTB 42 in \cite{boumisptb42} reports a H$\alpha$ flux of $12.7\times10^{-16}$ergs s$^{-1}$ cm$^{-2}$ arcsec$^{-2}$ \protect\AA$^{-1}$ and a logarithmic extinction at H$\beta$ of 1.41$\pm$0.04 mag. 
Investigating the evolution of the line ratios between the 2002 and 2018 spectra we see a decrease in nebular [O~{\sc iii}] emission with respect to H$\beta$, see Table \ref{lineratios2}. 
 
\begin{table*}
\centering
\caption{Comparison of the logarithm of PTB 42 line ratios, relative to the H$\beta$ line strength for the respective epoch, between the 2002 and 2018 spectra.} %Model (2018) corresponds to the best fit model from the grid described in Section \ref{pycloudy}, Fig. \ref{fig:pycloudy1}, with a shell density of 6.6 dex and a blackbody temperature of $1.4\times10^{5}$K, (although for the pyCross model an average density of 6.7 dex was assumed and a source temperature of $1.2\times10^{5}$K) - giving a marginally improved fit to the small number of emission line ratios.}
\label{lineratios2}
\begin{tabular}{cccccc}
\toprule
Object    & H$\beta$  & [O~{\sc iii}] 5007\protect\AA &  [N~{\sc ii}] 6583\protect\AA \\
\hline
PTB 42 (2002) & 0   & 1.5 & 1.2  \\
PTB 42 (2018) & 0   & 0.4 & 1.2  \\
\hline
%Model (2018) & 0  & 0.6 & 1.1 \\
\end{tabular}
\end{table*}
 
From the continuum subtracted 2016 imaging observations we find in the H$\alpha$ + [N~{\sc ii}] narrow-band exposure dimensions of 6.4$^{\prime\prime}$$\times$7.1$^{\prime\prime}$ 
for the nova shell, giving an uncorrected axial ratio of 
1.11. The [O~{\sc iii}] 
Aristarchos 2016 image gives 5.2$^{\prime\prime}$$\times$5.5$^{\prime\prime}$ 
and thus an axial ratio 1.06 (see Fig. \ref{fig:ptb42_imag} and Table 
\ref{tab:distmeas}), implying an increase in nebular diameter of 
0.32$^{\prime\prime}$/year. 
Extension measurements were taken from where the shell flux was 10\% above the background level in the (H$\alpha$ + [N~{\sc ii}]) - R band and ([O~{\sc iii}]) - B band images. 

Following \cite{2002AIPC..637..497B}, we use the inclination corrected axial ratio for the similar novae DQ Her, and T Aur which are given to be 1.4. Using this value find a probable inclination of the shell, and thus binary, to be $70 - 80^{\circ}$, 
consistent with the eclipse light curve seen in Fig. \ref{fig:PTB42brise2}.
However, considering an axial ratio of 1.6, as found for HR Del in \cite{HRDel3d} would imply a lower inclination for PTB 42, which would imply higher polar expansion velocities. This could explain the shallowness of the PTB 42 eclipse. To properly determine the axial ratio and inclination more detailed observations are required, such as the GMOS-IFU observations of HR Del in \cite{HRDel3d}.

%\begin{landscape}
%\end{landscape}

%\begin{landscape}
 \begin{figure*}
\centering
%\resizebox{{\columnwidth}{!}{
\includegraphics[width=14cm]{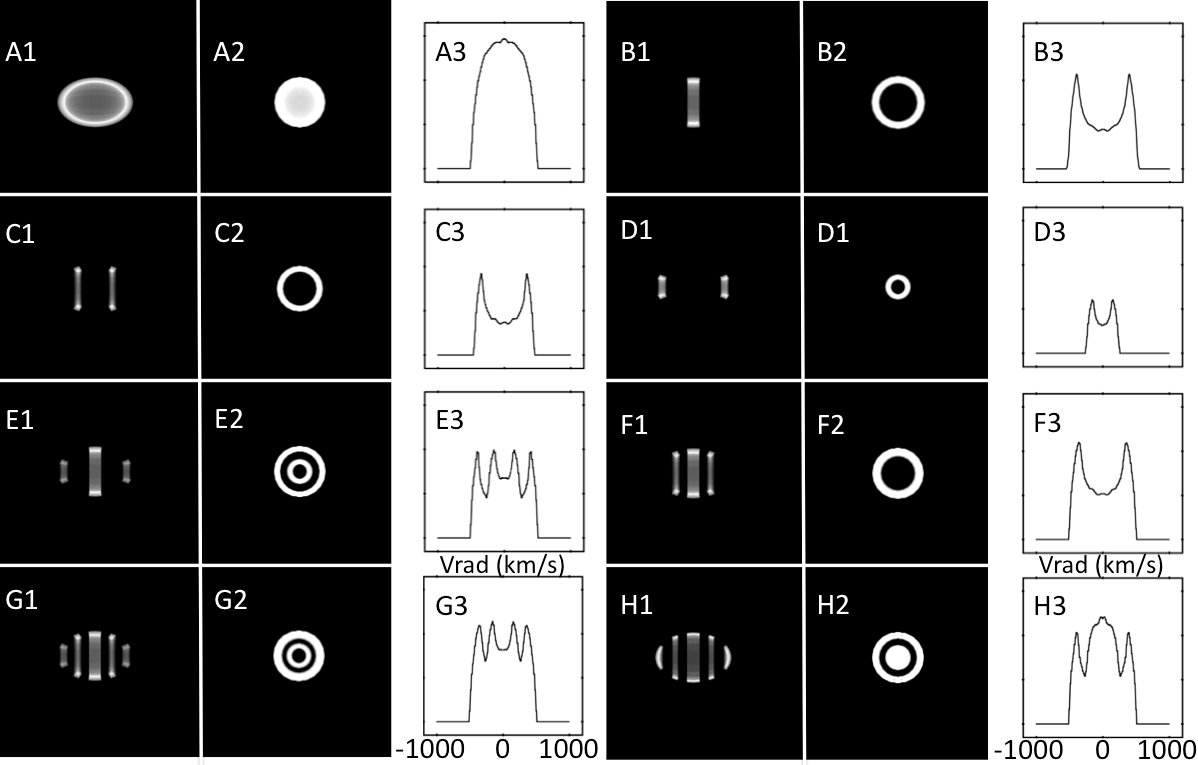}
\caption{$\textsc{shape}$ models of possible shell morphologies, all placed at an inclination of 90$^{\circ}$. Each letter (A-H) represents different morphologies: A) Elliptical shell; 
B) Equatorial waist only; C) Tropical rings only; D) Polar cones only; E) Equatorial waist + Polar cones; F) Equatorial waist and tropical rings; G) Equatorial waist, Tropical rings and Polar cones; H) Equatorial waist, Tropical rings and Polar blobs. Then the 1, 2 and 3 next to each letter depict the simulated 2D image, the resultant position-velocity array and the corresponding flattened 1D spectral line profile respectively.} 
\label{fig:linepossa}
\end{figure*}

%\begin{landscape}
 \begin{figure*}
\centering
%\resizebox{{\columnwidth}{!}{
\includegraphics[width=14cm]{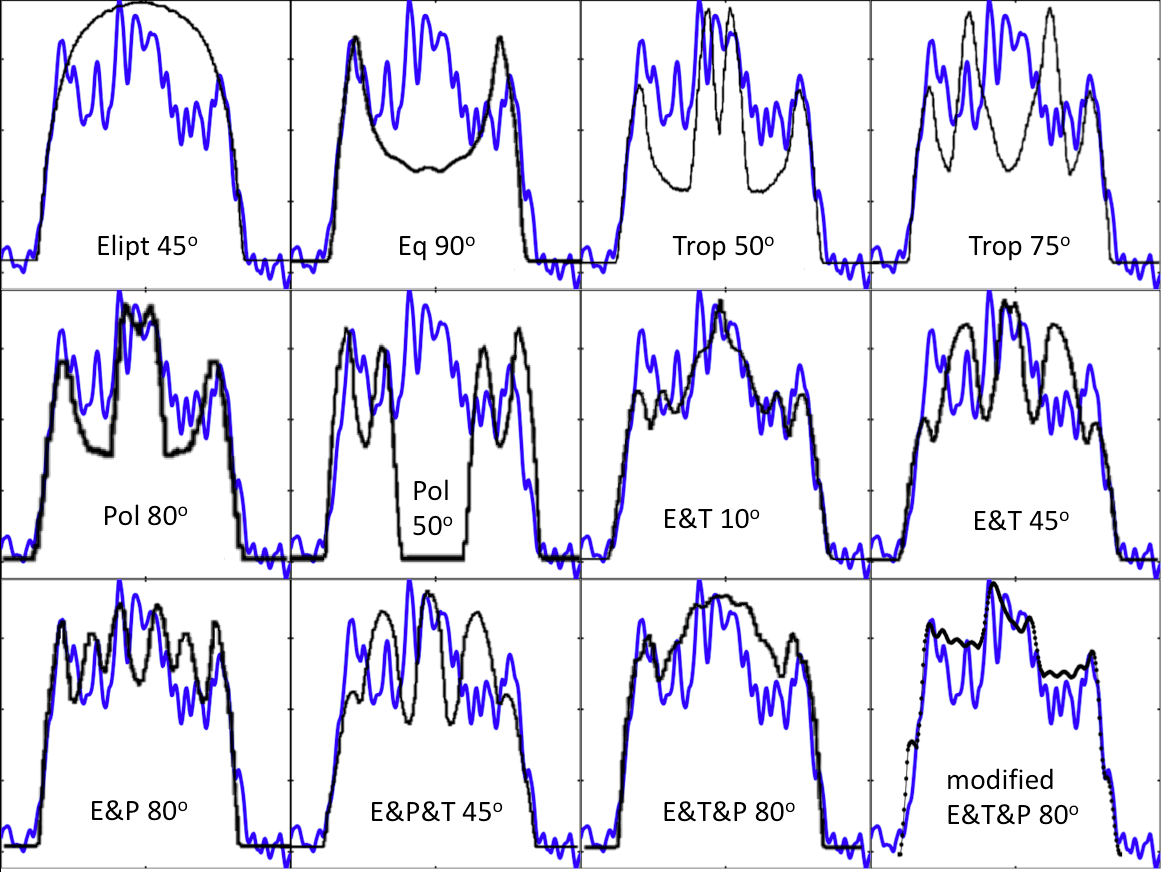}
\caption{$\textsc{shape}$ spectral line profile simulations - testing possible geometries to fit with the observed line profile. The tested geometries are illustrated in Fig. \ref{fig:linepossa}, and are overplotted in black here at their best fit inclination according to geometry. The blue line in each panel is the 2012 [N~{\sc ii}] observation of Fig. \ref{fig:ptb42_spec}. This analysis suggests that the observed spectral line features could be roughly reproduced by other morphologies. However, 2D line arrays (position-velocity arrays) of nova shells in the literature consistently suggest nova shells occupy a larger covering factor of an elliptical shell base than can be found with a two ring model (be they polar or tropical). Such that, by applying a double ring shell model we would not be consistent with the current known nova shell population. For example, the next best fit model in the figure, i.e. the polar cone model at an inclination of 80$^{\circ}$, is inconsistent with the narrow-band imaging for a structure only at the poles. Therefore, an equatorial waist, tropical ring and polar cone morphology is used for the final fit in Fig. \ref{fig:ptb42_spec}. The abbreviations in the plots are Elipt = filled eliptical shell; Eq = E = Equatorial ring; Trop = T = Tropical rings; Pol = P = Polar features. It is noted that shell morphology of novae is still an open debate and here we present simply our $\lq$best guess' given the observables at hand. High spatial reolution IFU spectroscopy is needed to properly untangle these structures.}
\label{fig:linepossb}
\end{figure*}

%\begin{landscape}
 \begin{figure*}
%\centering
%\resizebox{{\columnwidth}{!}{
\includegraphics[width=16cm]{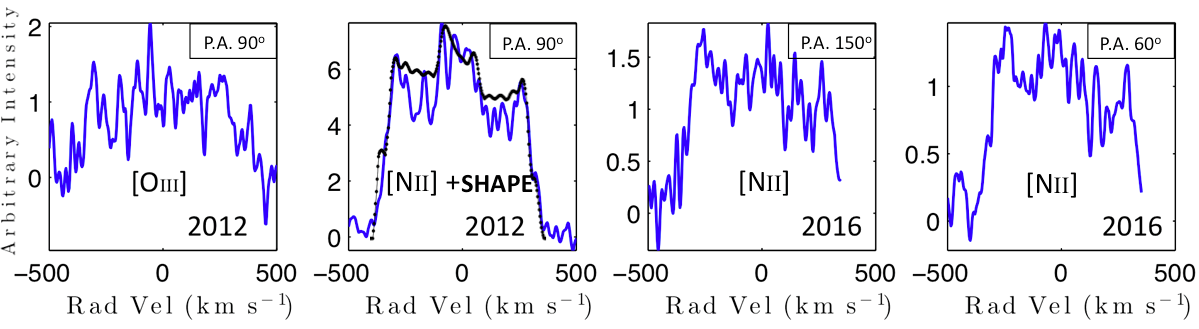}
\caption{PTB 42 [O~{\sc iii}] line and H$\alpha$ from MES observations taken in 2012 and 2016. Due to the shell becoming fainter over time the 2012 observations have higher S/N and higher spectral resolution (as a more narrow slit could be used). The two 2016 observations illustrated in the right-hand-side panels are from two different slit P.A (top right of each subplot). The repeated shape of the [N~{\sc ii}] line profile in the lower S/N 2016 observations, in comparison to the 2012 [N~{\sc ii}] observation, suggests that the $\lq$brighter blue-side' is real and likely due to the dust shell obscuring the nova shell's far-side. With low S/N in the [O~{\sc iii}] observation no dominant emission region can be identified in these observations. The [O~{\sc iii}] emission was undetectable using the instrument setup in 2016 due to the rapid fading of the shell.}
\label{fig:ptb42_spec}
\end{figure*}
%\end{landscape}

\begin{figure}
%\centering
%\resizebox{{\columnwidth}{!}{
\includegraphics[width=\columnwidth]{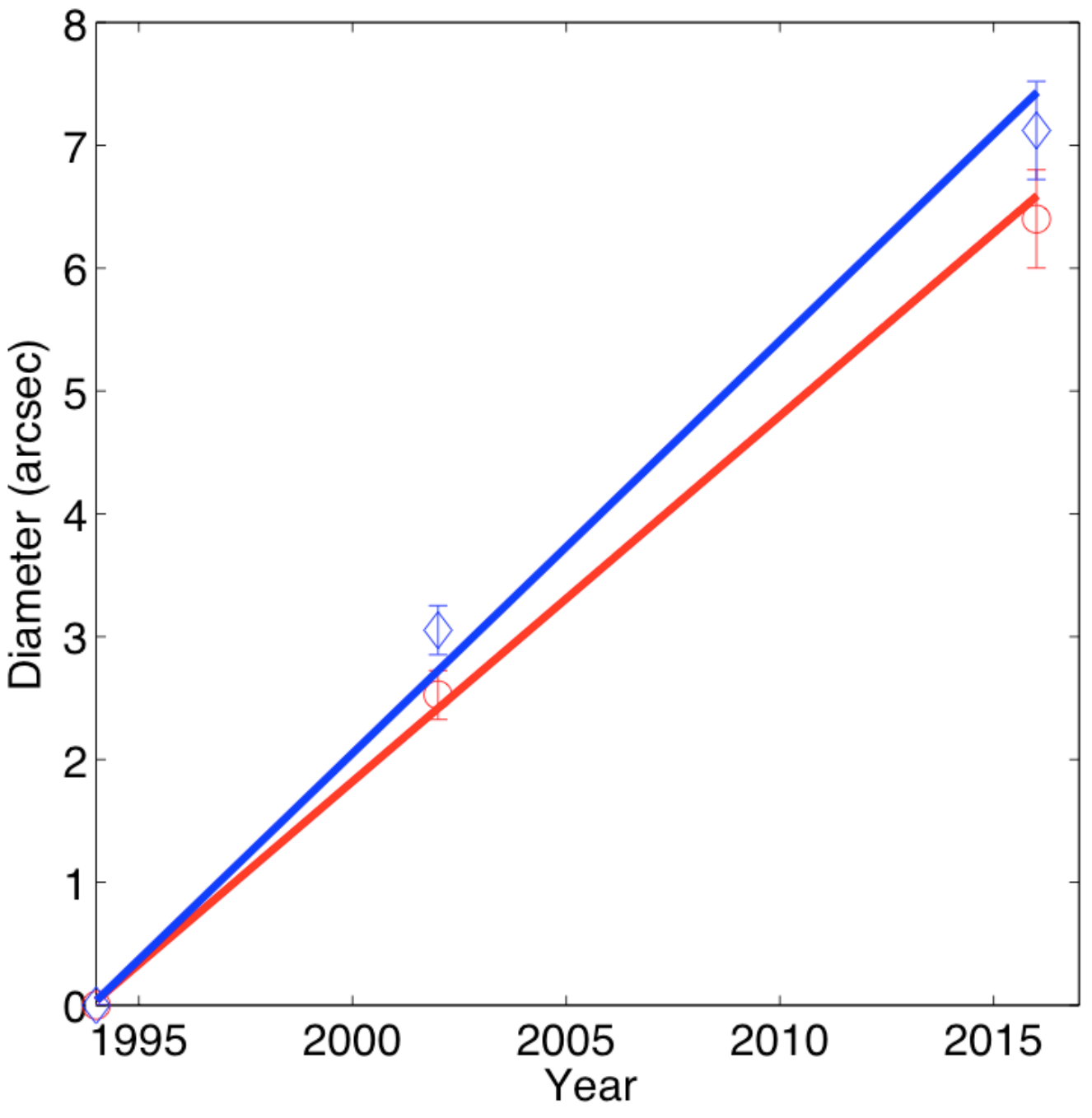}
\caption{Measured expansion from narrow band imaging, including uncertainty in measurement, of the PTB 42 shell surrounding the nova position of V4362 Sgr. Major axis represented by the blue line and minor axis by red. The shell is measured to expand at a rate of 0.32$^{\prime\prime}$/year.}
\label{fig:ptbexpand}
\end{figure}

Previously unpublished MES spectra from 2012 and newer 2016 spectra show a low velocity system, see Fig. 
\ref{fig:ptb42_spec} with suggestions of structure matching the description of \cite{evanspol}. 
The distance derived to this object (see Section \ref{distsect}) implies that the nova system is affected by interstellar reddening (otherwise the maximum absolute magnitude is of the order of -1). Or else, as this was a poorly observed nova the maximum of this ``erratic" nova was missed \citep{evanspol}. The system is known to be affected by circumstellar reddening with \cite{boumisptb42} calculating E(B-V)$_{obs}$ = 0.98, thus giving A$_{v}$ = 3.14, higher than the catalogued values of A$_{v}$ = 2.17 mag \citep{SFD} and A$_{v}$ = 1.87 mag \citep{SNF}. This is suggestive of local reddening at the source, that could be related to its dust shell. Although if related to the dust shell, formed later, then this would not have affected the observed peak magnitude. \cite{evanspol} were able to show that the observed polarisation signal was consistent with the presence of small dust grains. Looking at the NIR flux of the system (see Table \ref{tab:WISE_V4362sgr}) in the WISE archive \citep{WISE} and following the analysis prescription in \cite{wisenova} suggests the survival of a dust shell. This dust shell is expected to be the source of the asymmetry in the 2012 MES [N~{\sc ii}] line profile of Fig. \ref{fig:ptb42_spec}.

\begin{figure*}
%\centering
%\resizebox{{\columnwidth}{!}{
\includegraphics[width=17cm]{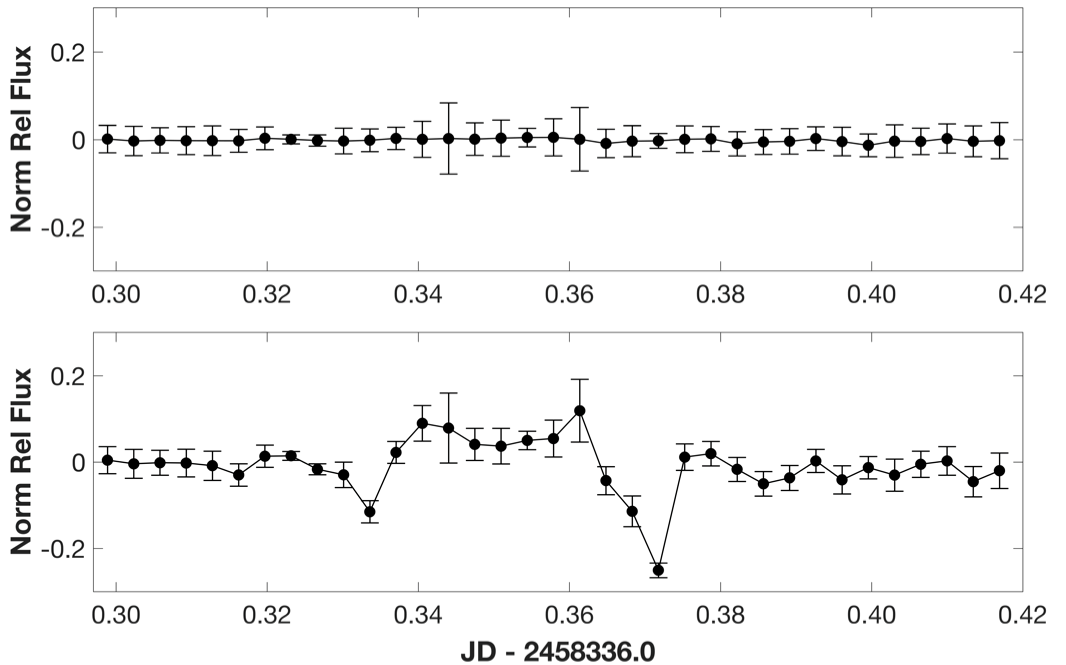}
\caption{In this figure the top panel shows the normalised averaged relative photometry of 11 non-variable field stars in the PTB 42 field of observation. The bottom panel is the corresponding PTB 42 light curve, on 5 August 2018. Observing using the RISE2~\protect\citep{rise2} instrument mounted on the Aristarchos telescope. 
Evidence of an eclipse can be seen, although higher time resolution observations are required in order to better constrain the system. The light curve shape is reminiscent of the eclipsing dwarf nova system IP Peg, see~\protect\cite{shafter_2neweclipse}. Suggesting the system to be eclipsing and therefore likely viewed at high inclination.}
\label{fig:PTB42brise2}
\end{figure*}

From the information gathered on the nova shell it is possible to start to build a 3D model. The polarimetry of 
\cite{evanspol} suggests a P.A. of around 150$^{\circ}$. 
This value was initially adopted for the P.A. of the shell. However, from examining the narrow-band images a P.A. roughly perpendicular to the P.A. of \cite{evanspol} is suggested after a polar cone opening angle is taken into account (110$^{\circ}$ difference i.e. of 40$^{\circ}$). This suggests that the dust polarisation observed in \cite{evanspol} arose from a ring-like equatorial structure (although \cite{evanspol} also found two perpendicular competing sources of polarisation). \cite{evanspol} put forward the idea that the observed broadband polarimetric behaviour in q, u space is symptomatic of a non-spherical and non-uniform shell, with an equatorial ring and polar blob shell structure being the favoured geometry. No inclination angle 
for the binary system exists that can be shown to be related to the inclination angle of the 
resultant nova shell, although the system is believed to be close to edge-on, i.e. 
of high inclination, supported by the eclipse light curve in Fig. \ref{fig:PTB42brise2}. It is noted here that the presence of an eclipse does not necessitate a high inclination, see for example T Aur \citep{bianchiniTaur} and some SW Sex stars, e.g. V795 Her in \cite{casares}. 

The common strong optical nebular lines ([O~{\sc iii}] and [N~{\sc ii}]) are visible in Fig. \ref{fig:spratptb42}. The rarefaction timescale found in \cite{warner} suggests density to decline as t$^{-3}$, where $\lq$t' is in weeks. Theory either suggests an initial density of 10$^{14}$ cm$^{-3}$ (early shocks), or an 
initial density of 10$^{10}$ cm$^{-3}$ (no-shocks), see \cite{derdzinski}. Assuming the presence of early shocks and considering the age of PTB 42 at 8114 days would suggest a shell density of  6.7 dex, i.e. in log(cm$^{-3}$).

\subsection{Object Distances}
\label{distsect}

\subsubsection{Gaia}
As neither nova was studied by \citet{schaefergaia}, the astrometrically derived distances presented here are based on the results table outlined in \cite{BailerJones2}. 

There has been recent discussion on the parallax offset in the Gaia DR 2 data release in relation to planetary nebulae (see 
\cite{stan2017}  and 
\cite{kimBerria2018}). Current \textit{Gaia} distances are affected by quiescent variability, nebulosity
and other influences of the orbit on observations, discussed in \citet{gaiawarn}. The systematic uncertainties surrounding
astrometric parallax for binaries will be better understood with the time-stamped \textit{Gaia} DR 4, until then it will be interesting to confirm
these distances by other methods.

Moreover, a systematic parallax offset in the \textit{Gaia} DR2  has been reported varying from 10 up to 100 mas, depending on the position of the sources in the sky, their magnitudes, and their colours, see \cite{2020MNRAS.492.4097G} with reference to \cite{bailerjones,kimBerria2018,2018ApJ...861..126R,2018A&A...619A...8G,2018MNRAS.481.1195M,2018ApJ...862...61S,2019ApJ...872...85G,2019A&A...625A.137S,2019MNRAS.489.2079L,2019ApJ...875..114X,2019MNRAS.486.3569H,2019ApJ...878..136Z}. For objects like novae and planetary nebulae, with compact and bluer central stars, the parallax offset has been properly estimated. The mean value from all the available measurements (0.051 mas) and the value derived from a sample of quasars (0.029 mas) were adopted for planetary nebulae by \cite{2020MNRAS.492.4097G}. The systematic uncertainties surrounding astrometric parallax for binaries will be better understood with the time-stamped \textit{Gaia} DR 4, until then it will be interesting to confirm
these distances by other methods.

It is noted in \cite{schaefergaia} that the \textit{Gaia} distance to DO Aql was not presented due to source confusion. On examination of the data there is a clear visual companion to DO Aql at an angular distance of 0.9", whereas the Gaia source detection limit is a separation of 0.3".

As the DO Aql source is known in this work from the association with a shell, its \textit{Gaia} source i.d. can be found in Table \ref{gaiaTab} along with that of PTB 42. However, for both DO Aql and PTB 42 their \textit{Gaia} parallax error exceeds the usable threshold given in \protect\cite{BailerJones2}.

For PTB 42 the nova progenitor system does not appear in the list of \citet{schaefergaia}. Since the distance found through the expansion parallax method is small for V4362 Sgr then why did it not become apparent on analysis of the nova population in the most recent \textit{Gaia} data release? 
There appear to be enough \textit{Gaia} visits to the source field to determine a reliable \textit{Gaia} parallax distance. Seemingly contrary to this the parallax significance is low.
%, possibly due to the strong extinction towards the source (100$\mu$m emission is 71.3$\pm$1.2 MJy sr$^{-1}$ in comparison to 1.8$\pm$0.1 for DQ Her). 
This motivates finding the \textit{Gaia} photometric excess factor, which turns out to be near the reliability threshold. 
Taking into account the cautionary notes for determining \textit{Gaia} distances to close binary systems without time-stamped data in \citet{gaiawarn}, as well as systems enshrouded in nebulae \citep{kimBerria2018,2019A&A...625A.137S} we explored the expansion parallax derived distance to PTB 42. Through the possession of knowledge of the P.A. of PTB 42 on the plane of the sky from comparing the broadband polarimetric observations of \citet{evanspol} and narrow-band imaging, the binary P.A. aligns close to the ecliptic. This effect will maximise the error of parallactic distance measurement, until the phase resolved (time-stamped) \textit{Gaia} release (DR4), discussed in \cite{gaiawarn}.

\begin{table*}
\centering
\caption{\textit{Gaia} \protect\cite{BailerJones2} distance estimates for both objects in this study. Distance units are parsecs.}
\label{gaiaTab}
\begin{tabular}{lllll}
\toprule

Obj    & Source i.d.         & modal     & -1$\sigma$     & +1$\sigma$     \\\hline
DO Aql & 4208116120913290752 & 3222.6611 & 1863.8114 & 5729.8343 \\
PTB 42 & 4096752394935572224 & 2211.0488 & 1291.2170 & 4756.0270

\end{tabular}
\end{table*}

\subsubsection{Expansion Parallax}

The expansion parallax method requires the astronomer to distinguish between nebular components, as misattributing components to velocities leads to errors in distance estimates \citep{wade2000}. In order to act conservatively, as the shells are poorly resolved in the discovery data, a relatively large error was derived by assuming that we cannot distinguish between equatorial and polar shell directions. 

\protect\cite{wade2000} and \protect\cite{porterasphericity} highlight effects of bipolarity/asphericity in the determination of distances to nova systems. Recently \protect\cite{2020ApJ...892...60S} studied the angular expansion of nova shells with respect to both equatorial and polar components and compared their results to \protect\cite{eamonnthesis}.

Distance measurements are shown for each of the major and 
 minor axes and epochs in Table \protect\ref{tab:distmeas}. 
 Final distances and errors are the mean of the distances measured from each axis and epoch.

\subsubsection{DO Aql}

The seeing in the 2017 observations presented was too poor to reliably measure the proper motion growth between it and the 2015 observation of the expanding shell. This is partially due to the large distance to the source and the small angular growth expected over 2 out of 92 years. However, both 2015 and 2017 observations have been used in the distance determination. The major axis is readily confirmed looking at Fig. 
\ref{fig:DOAql}, as well as the existence of small protrusions at both tips of the major axis that could be related to the ablated flows in the shell of HR Del, see \cite{Vaytet}. However, the minor axis is of a similar extension to the field stars and as such its measurement is uncertain.
The position angle (P.A.) of the nova shell, as measured from the H$\alpha$+[N~{\sc ii}] images, is taken to be 98$^{\circ}$ east of north. 

There are no previous distance estimates to the nova. Using the expansion parallax method a distance of 6.7 $\pm$ 3.5 kpc is found from measurements taken from the 2015 observations, (for a shell expansion velocity of 1000 km s$^{-1}$ , as reported in \cite{DOAqlVorontov}
), see Tables \ref{observationstab} and \ref{tab:distmeas}. Note that distance scales linearly with expansion velocity, as the shell velocity is poorly constrained for this nova we write the distance as 6.7 $\times (\mathrm{V_{exp}} / 1000 \mathrm{ km s^{-1}}) \pm$ 3.5 kpc. The distance cannot be confirmed through comparison with \textit{Gaia}. The object was not included in \citet{schaefergaia} due to source confusion. As the source is known for this work the \textit{Gaia} distance was checked in the context of the \citet{BailerJones2} method, see Table \ref{gaiaTab}. However, as the error in parallax is double the magnitude of the derived parallactic distance, the \textit{Gaia} measurements cannot be used to reliably determine a distance to DO Aql, see Section (\ref{discussion} for a discussion) and Table \ref{Inputparametersformodels}.

\cite{DOAqlVorontov} describes to the reader a spectrum taken 113 days post discovery by \cite{Merrill} on the Hooker 100$'$ that shows DO Aql to 
be an Fe~{\sc ii} type nova, with line ratios similar to those of V5668 Sgr at the same time post detection \citep{me_V5668Sgr}. \cite{DOAqlVorontov} measured a Balmer line expansion of 1000 km s$^{-1}$. In the low resolution SPRAT spectra acquired for this work a FWHM of 1330 - 1400 km s$^{-1}$ and a FWZI of around 2900 km s$^{-1}$ are determined from Balmer lines. However as no high-resolution spectrum was acquired for DO Aql, the expansion velocity determined from \cite{DOAqlVorontov} is used here to guide the distance calculation and the resultant uncertainty from minor and major axis distance determinations factored into the final error, see Table \ref{tab:distmeas}.

\subsubsection{PTB 42}

Using the expansion parallax relation presented in \cite{warner} and a measured shell expansion of 350 km s$^{-1}$ a distance of $0.5\substack{+1.4 \\ -0.2}$\ kpc is found, with the large relative error due to assigning the expansion to either the minor or major axes. As the source is eclipsing the nova is more likely to be on the closer end of the quoted distance scale. Which would make PTB 42 one of the closest and brightest known nova shells, see Table \ref{tab:distmeas}.

\begin{table*}[t]
\centering
\caption{Distance measurements according to measured major and minor axis diameters for the narrow band images of both nova shells. Calculated errors are a function of the seeing, uncertainty in expansion velocity (V$_{exp}$ in km s$^{-1}$), and scatter in the distance measurements according to the different distances suggested by the calculations for the stated epochs and filters.  The average errors and distances for both nova shells are in the final column. D represents distance from expansion parallax and are stated in kpc. Age is in days (d). Comparing the DO Aql [O~{\sc iii}] shell size with the SPRAT slit width it becomes evident why [O~{\sc iii}] was not observed, i.e. as the shell spectrum is extracted in the region flanking the stellar spectrum [O~{\sc iii}] was probably lost in the stellar spectrum. Also, as the [O~{\sc iii}] DO Aql shell is probably not associated with the outermost ejecta and was observed in poor seeing conditions it is not considered in its distance determination.}
\label{tab:distmeas}
\begin{tabular}{lllllllllll}
\toprule
Object & Age & Filter & Maj axis & Min axis & Seeing & V$_{exp}$ & D maj& D min  & err & D avg  \\
& (days) &  & axis & axis &  & (km s$^{-1}$) & (kpc) & (kpc) & & (kpc) \\
\hline
PTB 42 & 2928 & H$\alpha$ + [N~{\sc ii}] & 3.1$^{\prime\prime}$ & 2.5$^{\prime\prime}$ & 1.6$^{\prime\prime}$ & 350 & 0.37 & 0.46 & $\pm$1.51 &  \\
PTB 42 & 8114 & H$\alpha$ + [N~{\sc ii}] & 7.1$^{\prime\prime}$ & 6.4$^{\prime\prime}$ & 1.3$^{\prime\prime}$ & 350 & 0.45 & 0.50 & $\pm$1.38 &  \\
PTB 42 & 8114 & [O~{\sc iii}] & 5.5$^{\prime\prime}$ & 5.2$^{\prime\prime}$ & 1.2$^{\prime\prime}$ & 350 & 0.59 & 0.62 & $\pm$1.34 & $0.5\substack{+1.4 \\ -0.2}$\ \\
DO Aql & 32846 & H$\alpha$ + [N~{\sc ii}] & 6.6$^{\prime\prime}$ & 4.8$^{\prime\prime}$ & 1.8$^{\prime\prime}$ & 1000 & 5.6 & 7.8 & $\pm$3.36 &  \\
DO Aql & 33552 & H$\alpha$ + [N~{\sc ii}] & 6.6$^{\prime\prime}$ & 4.9$^{\prime\prime}$ & 2.3$^{\prime\prime}$ & 1000 & 5.8 & 7.8 & $\pm$3.52 & 6.7$\times(\frac{\mathrm{V_{exp}}} {1000 \mathrm{ km s^{-1}}})\pm$ 3.5\\
DO Aql & 33552 & [O~{\sc iii}] & 4.5$^{\prime\prime}$ & 3.3$^{\prime\prime}$ & 2.5$^{\prime\prime}$ & 1000 & &  &  &  \\
\end{tabular}
\end{table*}

\section{Simulations}
\label{mod}

In order to begin to build a 3D model of the PTB 42 nova shell its structure must first be untangled through interpretation of the observations. 
A similar technique to the following methodology was outlined in \cite{me_V5668Sgr}.
The P.A. is informed by polarimetric observations, in this case the study of \cite{evanspol}, as well as from close examination of the major and minor axes in the narrow-band imaging, see Fig. \ref{fig:ptb42_imag}. The high-resolution MES spectroscopy is then used to find the radial velocity of individual components of the nova shell. The source inclination is the most difficult value to derive, aside from the filling and covering factors. In order to arrive at an answer for the inclination, assumptions must be made, which are based on the system's expansion velocity by considering the shape of the individual spectral line profiles (and through study of the orbital signature of the quiescent light curve). However, the system inclination can be informed by the orientation of the equatorial ring \citep{slavin}. Complicating the situation is local reddening of the system, as can be seen most clearly in the shape of the [N~{\sc ii}] line in the 2012 observation in Fig. \ref{fig:ptb42_spec}, as well as the WISE observations summarised in Table \ref{tab:WISE_V4362sgr}.

Early photoionisation simulations of nova shells demonstrated the presence of possibly counter-intuitive phenomenology, such as the very low temperature of older nova shells \citep{1984ApJ...281..194F}. Novae tend to have enhanced C, N and O in comparison to solar abundances, although for some other novae they have been shown to have near solar abundances \citep{notallphoto1}. More recent work suggests that nova shells are not completely photoionised, but may also experience contributions from shock ionisation \citep{notallphoto2}.

After \cite{1984ApJ...281..194F}, efforts followed to understand the temperature and ionisation structure of nova shells \citep{Beck1}, as well as the effect of improving the radiation field \citep{Beck2}. A large body of work was to continue on interpreting and analysing nova spectra within the understood framework, see \cite{Vanlan}, \cite{shoreman}, \cite{spectrophot}, \cite{shoreRaman} and \cite{Mason_2018}.

To manage condensations and more complex structures associated with nova shells there are several available 3D or pseudo 3D codes available, notably RAINY3D \citep{RAINY3D,HRDel3d}, pyCloudy \citep{pycludy}, pyCROSS \citep{2020A&C....3200382F} and MOCASSIN \citep{mocassin}. 
%}

%\begin{landscape}
 \begin{figure}
\centering
%\resizebox{{\columnwidth}{!}{
\includegraphics[width=\columnwidth]{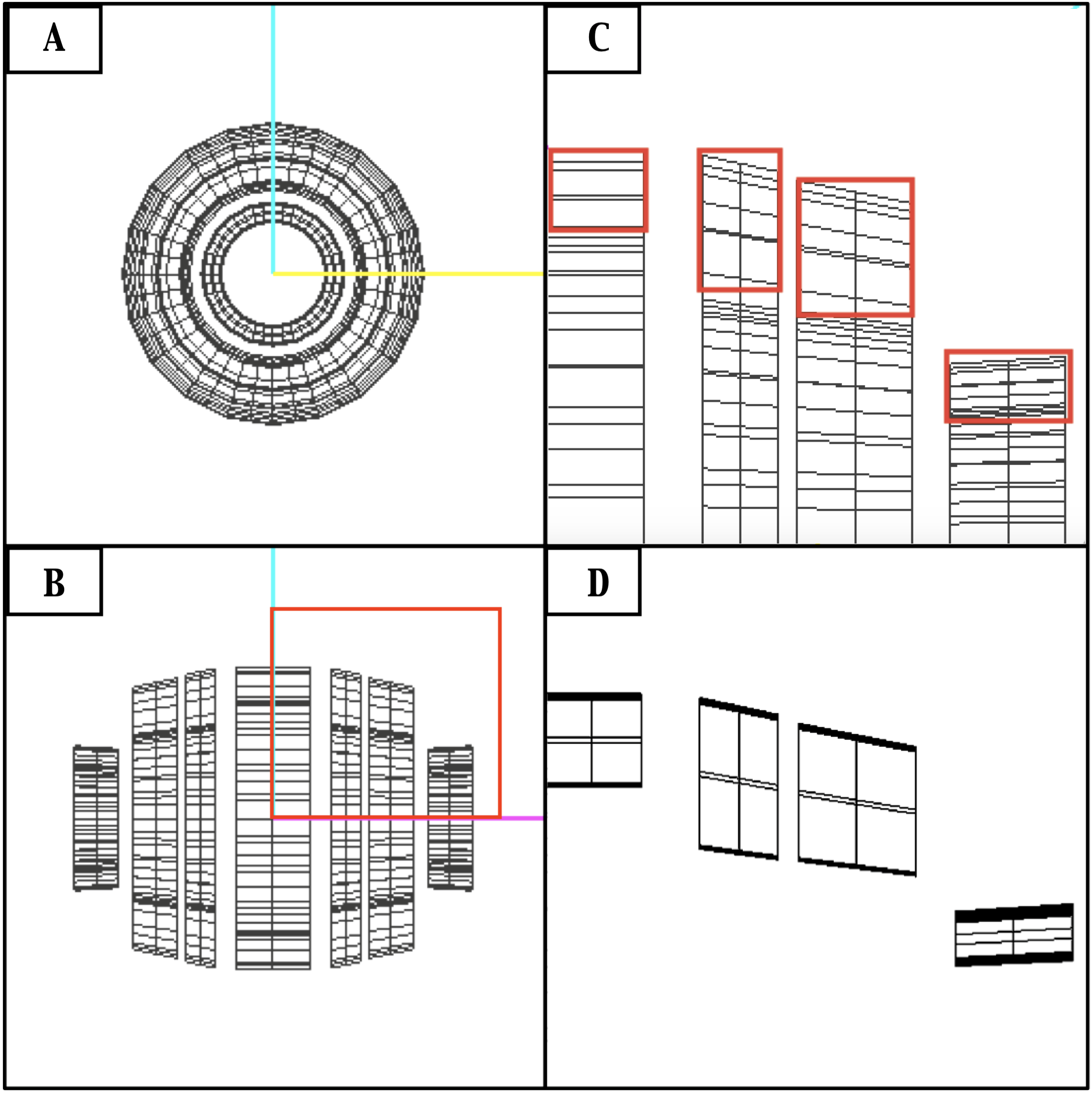}
\caption{$\textsc{shape}$ model used for input into $\textsc{pyCloudy}$, this figure shows the physical rendering of the fitted line profile in Fig. \ref{fig:ptb42_spec}. The top-left panel (A) shows the derived morphology of the nova shell as viewed pole-on, whereas the bottom-left panel (B) shows the structure as viewed edge-on. The top-right panel (C) is a zoomed in quarter of panel (B), i.e. the section with the red square. The 2D slice of the structure (seen in the bottom panel D and highlighted in red in panel (C) is output as a datacube and fed into $\textsc{pyCloudy}$ using a text file that describes the velocity, and density at each position in the shell. A set of 1D $\textsc{Cloudy}$ simulations are run through the 2D parameter space and are then wrapped in azimuth around the complete shell creating a pseudo 3D model.}
\label{fig:mesh}
\end{figure}

\begin{figure*}
\centering
%\resizebox{{\columnwidth}{!}{
\includegraphics[width=14cm]{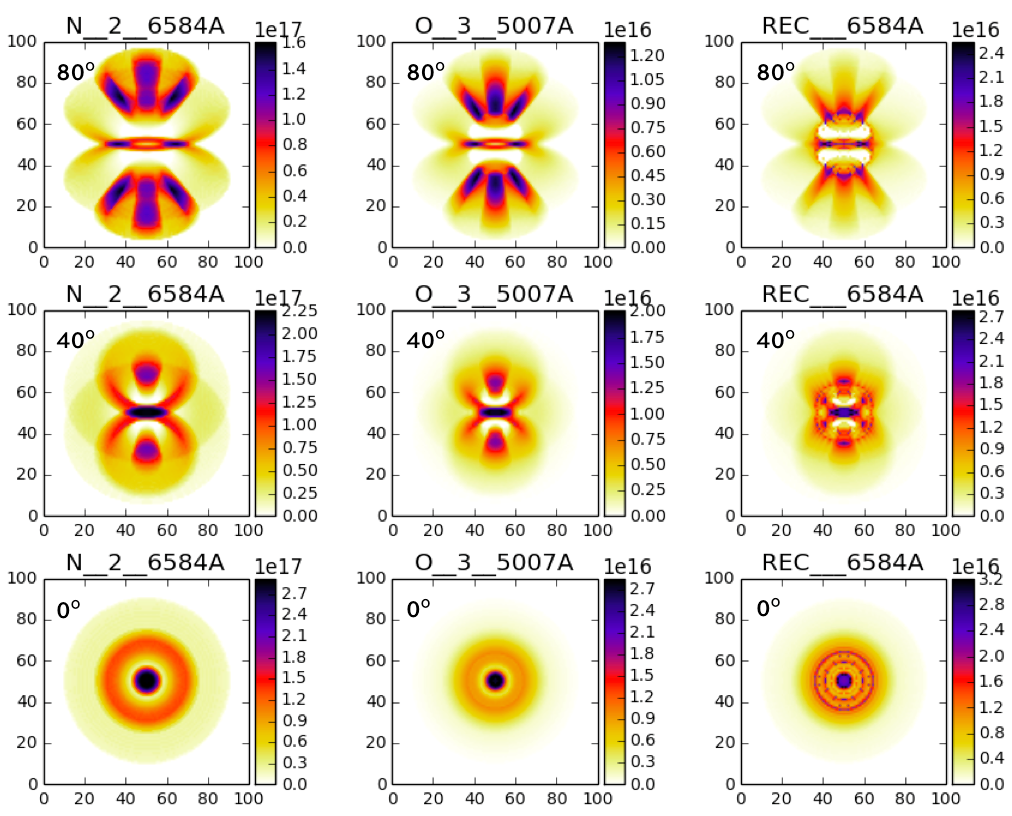}
\caption{$\textsc{pyCross}$ model of PTB 42. The x and y axes are in normalised distance units. The colour bar represents relative arbitrary flux for the associated line emission maps. Using DQ Her shell abundances from \protect\cite{1984ApJ...281..194F} and a luminosity of $\protect\times10\protect^{38}$ ergs s$\protect^{-1}$. From left to right the columns indicate the shell structure for ionisation processes involved in the production of the [N~{\sc ii}] 6584\protect\AA~line
combined in the first column. The second column shows the structure of the [O~{\sc iii}] 5007\protect\AA~emission line and the third column shows only the recombination structure of the [N~{\sc ii}] 6584\protect\AA~emission line. The top row is plotted at an inclination of
80$\protect^{\circ}$, middle row 40$\protect^{\circ}$ 
and bottom row is seen pole on at 0$\protect^{\circ}$ to demonstrate this effect for representative inclinations. As the system is likely an eclipsing binary, see Fig. \ref{fig:PTB42brise2}, the system is suspected to have an inclination 
$\protect\geq{60}\protect^{\protect\circ}$.
Note the [O~{\sc iii}] emission does not extend out as far as the [N~{\sc ii}] shell, as is often observed with respect to nova shells, e.g. GK Per \citep{GKme}. Although, the surface brightness is similar for the two emission lines, the ionisation of the more extended polar caps by the 
[N~{\sc ii}] shell may contribute to the larger observed line intensity of the source. If only the recombination component of the 
[N~{\sc ii}] 
6584\protect\AA~is considered then we would expect a similar extension and measured line strength to that of the [O~{\sc iii}]
5007\protect\AA~emission line.
}

%By including all commonly observed components of nova shells (i.e. equatorial ring, tropical rings and polar features) their individual behaviour can be examined here even if some of these features are found to be not present in future work.}
\label{fig:pycloudy1}
\end{figure*}

\subsection{SHAPE}
\label{shape}

A 3D morpho-kinematic $\textsc{shape}$ \citep{shapenew}\protect\footnote{\url{http://bufadora.astrosen.unam.mx/shape/}} model was created for the V4362 Sgr nova shell / PTB 42, see Figs. \ref{fig:ptb42_spec} \& \ref{fig:mesh}. Creating a full morpho-kinematic model of a poorly resolved nebula is non-trivial. Care must be made not to overintrepret limited observations with too many model elements. The spatial resolution constraints make it difficult to know the finer structure, i.e. the covering and filling factors related to the specific nova shell. Therefore, only the gross morphological parameters can be estimated from the observations presented here, i.e. the major and minor axis lengths. The P.A. can be estimated from polarimetric observations and/or narrow-band imaging, whereas the inclination requires knowledge of the binary system's orbital light curve or a fully resolved and distinguishable equatorial ring. Although the spatial information is not resolved, the structures can be resolved by line-of-sight velocities. If velocities along the plane of the sky are required they can be obtained through multi-epochal imaging.

To begin with various possible morphologies were tested through rotation around their inclination angle, see Figs. \ref{fig:linepossa} $\&$ \ref{fig:linepossb}. Following this a morphology consisting of an equatorial waist, tropical rings and polar cones was chosen. PTB 42 is thought to be at high inclination, therefore the highest observed velocities would be from the equatorial disk although if all velocities were deprojected the polar velocities would be expected to be higher. The observed equatorial velocity is 350 km s$^{-1}$, as measured from the MES spectra. Then, for an axial ratio of 1.4, i.e. the inclination corrected axial ratio for similar novae DQ Her and T Aur \citep{BodeNova}, gives a polar velocity of 490 km s$^{-1}$. Adjusting for inclination when fitting to the asymmetry in the line profile gives an equatorial velocity of 390 km s$^{-1}$ and polar velocity of 550 km s$^{-1}$. This allowed for the remaining velocities to be set to $550\times (\mathrm{{r}/{r_{0}}})$ (km s$^{-1}$). Looking at the line profiles of Fig. \ref{fig:ptb42_spec} the gross morphology of the castellated features are not noise as they are present in multi-epoch observations. The [N~{\sc ii}] line profile from the 2012 observation, seen in the second panel from the left in Fig. \ref{fig:ptb42_spec}, was chosen for modelling as it had the highest S/N (due to the shell becoming fainter at later times and has the best velocity resolution due to the more narrow slit used). Substructure in the line profiles could be due to the presence of clumps although on more narrow velocity scales, due to their relatively smaller individual sizes. This implies that the degree of clumping cannot be deduced from these observations. However, it informs that the observed gross structure is related to physically real features. As such, the components can be associated with polar blobs, equatorial waist and tropical rings in the $\textsc{shape}$ model, with the tropical rings suggested by the emission intermediate of the central peak and outer wings. As the system is suspected to be viewed at a high inclination, the broadest observed velocity features are expected to arise from the lower velocity equatorial waist.

Illustrated in Fig. \ref{fig:linepossa} are several morpho-kinematic models that demonstrate the relationship between image, position velocity (PV) array and 1D line spectrum for commonly proposed nova shell morphologies, all viewed at 90$^{\circ}$. In Fig. \ref{fig:linepossb} the best fit inclinations of the various models are shown plotted over the 2012 MES observation. Deep observations with a high resolution IFU spectrograph (or a long-slit spectrograph on an $\sim$ 8m class telescope) would allow to fully distinguish between the possible morphologies. It should be mentioned regarding the ionisation simulations of Fig. \ref{fig:pycloudy1} that the effects of the equatorial waist, tropical or polar components are independent such that they can be decoupled from each other.
The line profile shapes suggest PTB 42 is viewed at high inclination, which is also supported by the quiescent orbital light curve of the system presented for the first time in Fig. \ref{fig:PTB42brise2}. The quiescent light curve, although it requires better temporal sampling, is similar to known eclipsing systems such as T Aur \citep{walker_TAur,bianchiniTaur} and the dwarf nova IP Peg, see \cite{shafter_2neweclipse}. In order to match the observed asymmetry in the line profiles the morphology of the object was modified such that the flux contribution of the red-shifted portion of the shell was reduced by 5$\%$, see Fig. \ref{fig:mesh}. However, the [N~{\sc ii}] line asymmetry may be due to contamination by H$\alpha$, situated just blue-ward of the plotted [N~{\sc ii}] line. The density structure of the nova was assumed to be 6.7 dex, as suggested by the $\textsc{pyCloudy}$ \citep{pycludy} grid of Fig. \ref{fig:pycloudy01}, and in agreement with theoretical predictions that assume early interacting shocks \citep{derdzinski}.

\subsection{Shell Ionisation}
\label{pycloudy}

In an attempt to represent a snapshot of the PTB 42 shell $\textsc{pyCloudy}$ \citep{pycludy} was used in this work to both control $\textsc{Cloudy}$ \citep{cloudy} and interpret $\textsc{shape}$ output data. As PTB 42, as well as nova shells more broadly, are not solely photoionised, other sources of ionisation must be taken into account. The $\textsc{Cloudy}$ code takes collisional ionisation and recombination into account, as well as effects of turbulence. However, shock ionisation is not considered. Early stages in nova shell excitation arises from a number of processes (although thought to be mostly photoionisation from the UV bright white dwarf, shock ionisation also plays an important role during these early times) and at late times the shell enters a regime of pure recombination. The switch from 'early time' to 'late time' depends on the outburst characteristics on the nova event and can range from a few days for the fastest systems, up to years for the slowest evolving and expanding shells, see V1280 Sco as described in \cite{Chesneau:2012aa}. Fossil nova shells (such as that observed in M31N 2008-12a \citep{DarnleyM31shell}, as well as galactic examples V2275 Cyg \citep{sahman}, AT Cnc \citep{Shara:2012ac} and Z Cam \citep{Shara:2012ab}) are thought to be mostly shock excited.

The $\textsc{shape}$ model as determined from the PTB 42 line profile, Fig. \ref{fig:ptb42_spec} and Section \ref{shape}, can be output in a data cube, which in turn can be read by a modified version of $\textsc{pyCloudy}$. This pairing routine between $\textsc{shape}$ and $\textsc{pyCloudy}$ was first presented in \cite{me_V5668Sgr} and is referred to as $\lq\textsc{pyCross}'$. A detailed description of this code will be featured in \cite{2020A&C....3200382F}. 

\begin{figure*}
\centering
%\resizebox{{\columnwidth}{!}{
\includegraphics[width=10cm]{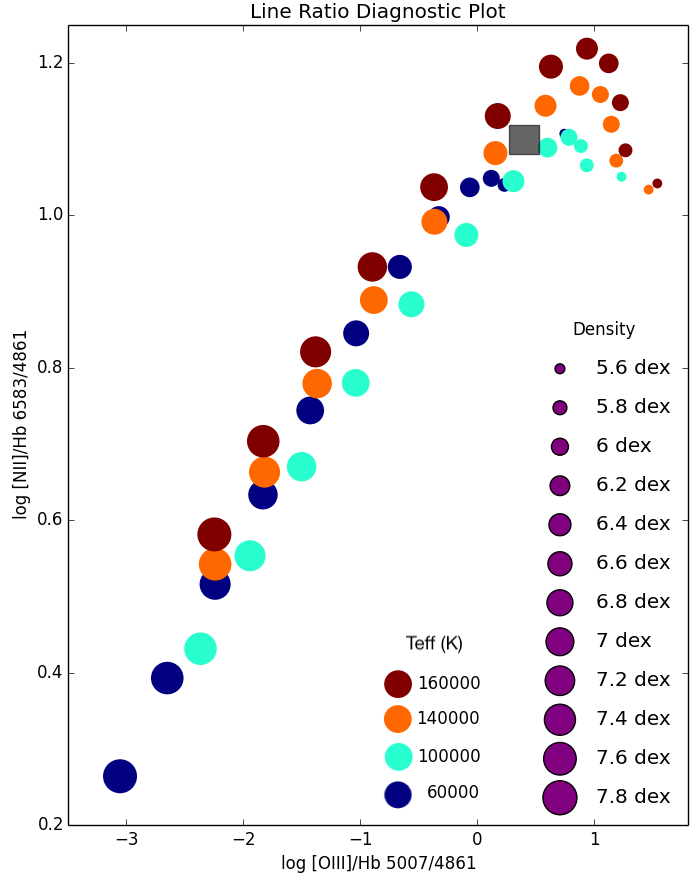}
\caption{$\textsc{pyCloudy}$ simulation grid for the PTB 42 nova shell. Using DQ Her abundances from \protect\cite{1984ApJ...281..194F}. The size of the blue square marks the observed measured line ratios 
from the SPRAT spectrum of PTB 42 discussed in Section \protect\ref{Spectroscopy}, and its size equivalent to the uncertainty in line ratio determination. The colour bar on the right provides a key to deciphering the effective temperature of the ionising blackbody. The plot suggests a high-density nova shell, on the order of 6.5-7.5 dex.}
\label{fig:pycloudy01}
\end{figure*}

\begin{figure*}
\centering
%\resizebox{{\columnwidth}{!}{
\includegraphics[width=17.5cm]{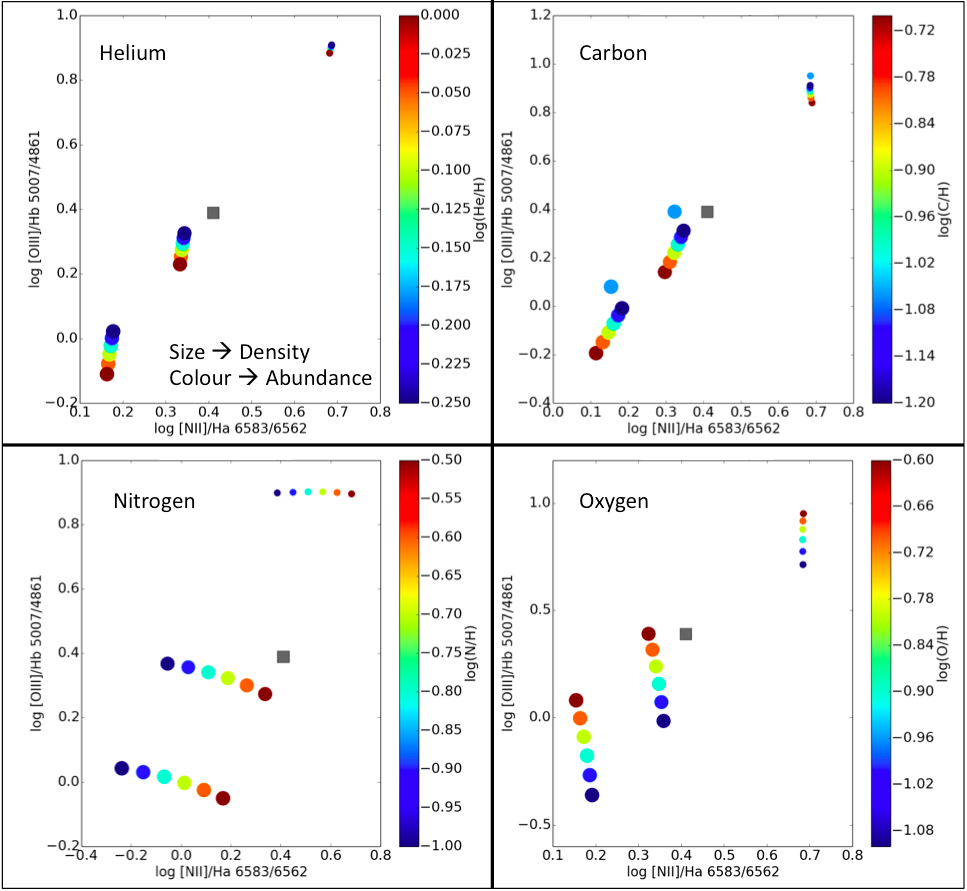}
\caption{$\textsc{pyCloudy}$ simulation grids for the PTB 42 nova shell. The suitability of DQ Her abundances from \protect\cite{1984ApJ...281..194F} are tested. The blue square marks the observed measured line ratios 
from the SPRAT spectrum of PTB 42 discussed in \protect\ref{Spectroscopy}. The colour bar on the right indicates the relative abundance of the element relative to Hydrogen. The plots suggests that relative to DQ Her nova shell abundances that helium and carbon are less abundant and nitrogen and oxygen more abundant. The squares signify the measured observed line ratios. The densities used in these model grids are 6.5, 7.25 and 7.5 dex (in all four panels the highest to lowest density values group from the bottom left corner to the top right corner). These model grids suggest that the helium abundance is more abundant in PTB 42 than in the nova shell of DQ Her, with the same being true for carbon and the opposite for nitrogen and oxygen. However, these suggestions should be taken with a large caution, such that at a different density the inverse interpretation can be arrived at. Therefore, we do not have enough observational constraints to identify the abundances of the nova shell.}
\label{fig:pycloudy03}
\end{figure*}

Before creating models, the conditions must first be understood. The luminosity of the system is based on the quiescent luminosity of DQ Her as was measured in \cite{1984ApJ...281..194F}. Archives were searched through for UV and X-ray observations of the object, targeted or serendipitous, however unfortunately there were none. The inner and outer radii of the shell are estimated based on the observed expansion velocity distribution and narrow-band imaging, although the actual shell thickness is difficult to know without resolving it spatially. Abundances of the archetypal slow nova, DQ Her, were used \citep{1984ApJ...281..194F}, although a later test is used to check the effect of this assumption, see Fig. \ref{fig:pycloudy03}. The free parameters that were iterated over are nebular density and central blackbody effective temperature until a satisfactory fit was reached. Line ratios estimated by individual models within the grid are extracted and plotted in Fig. \ref{fig:pycloudy01}. From this an estimate of the shell density and effective temperature can be found for any observable set of line ratios included in the database, at the distance estimated to the nova from us and the shell from the ionising source. Although the recovered spectral lines are reasonable density indicators, unfortunately they are not good temperature diagnostics at the high densities found here.

As $\textsc{pyCloudy}$ drives a 1D ionisation code, the 3D $\textsc{shape}$ model is simplified to 2D by taking a slice section of the $\textsc{shape}$ model (see bottom right panel of Fig. \ref{fig:mesh}). 
The spatial and velocity information is recorded in a data cube, which is then read by the modified version $\textsc{pyCloudy}$ and a series of 1D 
$\textsc{Cloudy}$ simulations are computed along the 2D slice $\textsc{shape}$ model. 
Then $\textsc{pyCloudy}$ wraps the 2D ionisation map around and flips it in order to create the full pseudo-3D photoionisation model, see Fig. \ref{fig:pycloudy1}. It is interesting to note that observed line ratios are also inclination dependent, further complicating the problem. This technique is constrained to axisymmetric nebulae. 

The grid of models presented in Fig. \ref{fig:pycloudy01} sample the density and blackbody temperature parameter space for the 1D $\textsc{pyCloudy}$ models. The free parameters were density (5.6-7.8 dex in 0.2 dex increments) and blackbody effective temperature (sampled at 60,000K, 100,000K, 140,000K and 160,000K). The fixed parameters were inner and outer shell radii (10$^{16.65}$ - 10$^{17.25}$ cm), DQ Her nova shell abundances from \cite{1984ApJ...281..194F}, shell age (22 years), turbulent velocity (300 km s$^{-1}$, affecting line width), source distance (600 parsecs) and a luminosity of $1\times10^{38}$ ergs s$^{-1}$ \citep{1984ApJ...281..194F}. The shell radius was informed by the observed expansion velocity and age of the nova. 

The pseudo 3D model was then generated with the parameter fit to the $\textsc{pyCloudy}$ grid as well as the geometry from the 
$\textsc{shape}$ model. Despite the number of assumptions required this basic model replicates the [N~{\sc ii}] and 
[O~{\sc iii}] emission distribution observed. The fit suggests a shell density in the range of 6.4 - 6.8 dex, see Fig. \ref{fig:pycloudy01}. A shell density of 6.6 dex and blackbody effective temperature of 10,000K give an average shell electron temperature of 5,800K.

However, as other sources of ionisation could not be simulated within the presented framework the ionisation source effective temperature is overestimated given the poor temperature dependence of the recovered emission lines, at the derived densities. As such, the effective temperature derived for these models cannot be used. Although the observed lines are dependent on shell density, under the conditions present in the shell. Line strengths in this model include recombination, collisional and photoionisation contributions. The code cannot simulate shock ionisation conditions.

\begin{table*}
%\centering
\caption{Results of measurements of the newly discovered nova shells surrounding the V4362 Sgr (PTB 42) and DO Aql nova positions. R$_{in}$ and R$_{out}$ represent inner and outer radii, respectively. The units for R$_{\mathrm{in}}$ and R$_{\mathrm{out}}$ are log(cm). Exp D represents the distance found via the expansion parallax method, and \textit{Gaia} D the \protect\cite{BailerJones2} distance.}
\label{Inputparametersformodels}
\begin{tabular}{llllllll}
\toprule
Object    & Max Date   & Model Date & Age (d) & Exp D (kpc)  & \textit{Gaia} D (Kpc) & R$_\mathrm{{in}}$        & R$_\mathrm{{out}}$         \\
\hline
PTB 42 & 16/05/1994 & 02/08/2016 & 8114   & $0.5\substack{+1.4 \\ -0.2}$     &   $2.2\substack{+2.5 \\ -0.92}$     & 16.45 & 16.85     \\
DO Aql   & 14/09/1925 & 19/08/2015 & 32846  &  $6.7\times (\frac{\mathrm{V_{exp}}} {1000 \mathrm{ km s^{-1}}}) \pm$ 3.5   & $3.2\substack{+2.5 \\ -1.4}$  &   17.2 & 17.36     
\end{tabular}
\end{table*}

To summarise the process employed: to understand ionisation conditions broadband spectra are required from which line ratios are measured, ideally including UV and NIR lines. Then $\textsc{pyCloudy}$ is used to run a grid of $\textsc{Cloudy}$ models, the best fitting model parameters are then run through the derived geometry. The geometry is found from matching line profiles in high-resolution spectra and narrow-band imaging in the $\textsc{shape}$ software. Polarimetry can be used to inform the P.A. of the shell. The inclination of the shell is related to the inclination of the binary, which can be found reliably if the binary system is eclipsing. With abundances adapted from the DQ Her nova shell model of \cite{1984ApJ...281..194F}, a pseudo 3D simulation of the ionisation structure of PTB 42 / V4362 Sgr is constructed and can be seen in Fig. \ref{fig:pycloudy1}. The results show the difference in emission regions for the strongest nebular lines, i.e. [N~{\sc ii}] and [O~{\sc iii}]. Although not shown, Balmer lines, for nebulae in general, trace the [N~{\sc ii}] emission.

 %\begin{figure}
%\centering
%\resizebox{{\columnwidth}{!}{
%\includegraphics[width=9cm]{06_07_2018.png}
%\caption{PTB 42 lightcurve. First night observing - 6/July/2018.}
%\label{fig:PTB 42-a-rise2}
%\end{figure}

\section{Discussion}
\label{discussion}

In this paper, two previously undiscovered classical nova shells are uncovered and an analysis is conducted in an attempt to decipher gross characteristics. The two nova shells surround nova systems of the DQ Her type. Unfortunately, they were both poorly observed during eruption and maximum magnitude possibly missed, although more applicable to V4362 Sgr. 

Due to fortuitous multi-epoch observations of the circumstellar environment of the two nova systems studied, distances were estimated. Distance estimation from the expansion parallax method are reliable and provide a good cross check for distances derived in the \textit{Gaia} era. Although both novae reported on here are bright and close enough to be recovered by \textit{Gaia} DR2 
\citep{gaiadr2}, on examining the \citet{BailerJones2} parallaxes and the results on the novae discussed in this work problems are present. Firstly, \citet{schaefergaia} reports that the distance to DO Aql could not be reported due to source confusion. Here, since the nova progenitor is identified through the associated shell the source can be identified, but since the parallax error is twice that of the measured parallax, distance measurements are not reliable. \textit{Gaia} data shows a visual companion separated from DO Aql by $\sim$0.9$^{\prime\prime}$. The objects can be distinguished via their colours, with DO Aql being the brighter bluer object. The \textit{Gaia} parallactic distance to DO Aql is thus $1.5\substack{+1.7 \\ -0.6}$ kpc, implying the expansion velocity reported by \cite{DOAqlVorontov} from the 1926 spectrum may be an overestimate. This would better explain why a shell is observed around the DO Aql system, as they are generally observed around nova systems within the nearest kpc or two.

From $\textsc{pyCloudy}$ simulations the long
term evolution of density conditions in nova outflow requires early interacting shocks to 
sustain an observable nova shell at late times, as suggested by \cite{derdzinski}. 
Higher densities in this way require a low filling factor to be consistent with the ejected shell mass estimates derived from radio observations of nova shells. A high degree of clumping is observed in most, if not all, nova shells resolved to the required degree to distinguish such phenomenology, see for example the well resolved shell of GK Per \citep{Seaquist89,anuandprabu93,Liimets:2012aa,Shara:2012aa,GKme}. 

The most apparent difficulties that arise during analysis are in deriving the opening angles of polar and equatorial features with respect to the central system as well as the degree of clumping. At very early, as well as at late times, shocks are expected to play a role in clumping and the ionisation of nova shells \citep{notallphoto1,notallphoto2,derdzinski}. 

%An idea that is gaining momentum of late \citep{derdzinski,notallphoto2}, although hinted at \textbf{over} the past few decades \citep{notallphoto1}, is that pure photoionisation simulations fall short of explaining the phenomonology witnessed in classical nova shells. This begs the question whether the illumination from these shells is purely due to photoionisation effects \textbf{and} recombination or whether shock excitation is partially responsible for the long term illumination of these enigmatic objects. 

%In this paper, the differences between the SPRAT DO Aql and PTB 42 spectra is that DO Aql's \textbf{shell spectrum is dominated by the stellar component}. As PTB 42 is much closer to us the shell is less affected by the progenitor PSF, as it appears relatively more detached from the parent system. Also, since the PTB 42 shell is brighter, as it is younger, its relative brightness in comparision to the progenitor system is greater. 

%\section{Conclusions}
%\label{conclusions}

Here it is shown that from limited multi-epoch data from small-medium sized research telescopes and archival data new nova shells can be revealed. This holds true even if the nova eruption was poorly observed, as is the case with the two novae studied here. PTB 42 appears to be one of the closest nova systems ever observed from work presented here. Possibly due to its difficult to observe position in the Northern summer sky the system had not been identified by the nova community as an object worth extensive follow-up. With both the discovered shells surrounding DQ Her-like nova systems, and both eclipsing, they are attractive for follow-up studies.

The potential number of undiscovered nova shells is large (with only 10$\%$ of the uncovered galactic nova population 
having been shown to harbour shells) with many more systems are now expected to have resolvable shells with large aperture or space-based telescopes.
As the nova shells inform the observer on aspects of the nova eruption and underlying binary, we 
finish with an appeal: for more such searches and follow up deep observations that aid in untangling the geometry, ionisation conditions and system abundances. To date nova studies have overwhelmingly focused on the nova event, but novae are non-destructive and their shells reveal information on the characteristics of the circumstellar medium, orientation of the underlying binary on the plane of the sky, the abundances of the secondary, the ejected mass, the white dwarf mass and chemical enrichement through thermonuclear burning processes.

%
%%%%%%%%%%%%%%%%%%%%%%%%%%%%%%%%%%%%%%%%%%%%%%%%%%%%%%%%%%%%%%%%%%%%%
% This work was supported by the following:
 
\section*{acknowledgements}

The authors would like to thank the staff at SPM and Helmos observatories for the excellent support received 
during observations. The Aristarchos telescope is operated on Helmos Observatory by the IAASARS of the National Observatory of Athens. The Liverpool Telescope is operated on the island of La Palma by Liverpool John Moores University in the Spanish Observatorio del Roque de los Muchachos of the Instituto de Astrofisica de Canarias with financial support from the UK Science and Technology Facilities Council.
MJD acknowledges support from the UK Science \& Technology Facilities Council (STFC). This publication makes use of data products from the Wide-field Infrared Survey Explorer, which is a joint project of the University 
of California, Los Angeles, and the Jet Propulsion Laboratory/California Institute of Technology, 
funded by the National Aeronautics and Space Administration. This project was funded in part by the Irish Research Council's postgraduate funding scheme. We also wish to acknowledge the databases used that made calculations in this work possible. Namely recombination coefficients were taken from $http:$$/$$/$$amdpp.phys.strath.ac.uk/tamoc/RR$ and  $http:$$/$$/$$amdpp.phys.strath.ac.uk/tamoc/DR/$ and the ionic emission data is from version 7.0 of CHIANTI. CHIANTI is a collaborative project involving the NRL (USA), the Universities of Florence (Italy) and Cambridge (UK), and George Mason University (USA).

% The best way to enter references is to use BibTeX:
\section*{Data Availability Statement}

All WISE data used in the preparation of this manuscript is available from \url{https://irsa.ipac.caltech.edu/Missions/wise.html}. Liverpool Telescope data can be acquired from \url{https://telescope.livjm.ac.uk/cgi-bin/lt_search} of both raw and calibrated datafiles. Imaging data from the Skinakas and Aristachos telescopes, as well as spectroscopy data from MES at the SPM will be available at \url{https://github.com/EJH-ljmu/TwoShells}.

\bibliographystyle{mnras}
\bibliography{nova_nova_references1} % if your bibtex file is called example.bib % if your bibtex file is called example.bib

% Alternatively you could enter them by hand, like this:
% This method is tedious and prone to error if you have lots of references
%\begin{thebibliography}{99}
%\bibitem[\protect\citeauthoryear{Author}{2012}]{Author2012}
%Author A.~N., 2013, Journal of Improbable Astronomy, 1, 1
%\bibitem[\protect\citeauthoryear{Others}{2013}]{Others2013}
%Others S., 2012, Journal of Interesting Stuff, 17, 198
%\end{thebibliography}

%%%%%%%%%%%%%%%%%%%%%%%%%%%%%%%%%%%%%%%%%%%%%%%%%%

%%%%%%%%%%%%%%%%% APPENDICES %%%%%%%%%%%%%%%%%%%%%

%%%%%%%%%%%%%%%%%%%%%%%%%%%%%%%%%%%%%%%%%%%%%%%%%%

% Don't change these lines
\bsp	% typesetting comment
\label{lastpage}
\end{document}